\documentclass[traditabstract]{aa}
\usepackage{graphicx}
\usepackage{txfonts}
\usepackage{epsfig}
\usepackage{natbib}
\usepackage{color}

\begin{document}

\title{Chemodynamical evolution of the Milky Way disk I: The solar vicinity}

\author{I.~Minchev\inst{1}, 
C.~Chiappini\inst{1}, 
M.~Martig\inst{2}
}

\institute{Leibniz-Institut f\"{ur} Astrophysik Potsdam (AIP), An der Sternwarte 16, D-14482, Potsdam, Germany
\email{iminchev1@gmail.com}
\and
Centre for Astrophysics \& Supercomputing, Swinburne University of Technology, P.O. Box 218, Hawthorn, VIC 3122, Australia}

\date{Accepted 1 July 2013}

\abstract{
In this first paper of this series, we present a new approach for studying the chemo-dynamical evolution in disk galaxies, which consists of fusing disk chemical evolution models with compatible numerical simulations of galactic disks. This method avoids known star formation and chemical enrichment problems encountered in simulations. Here we focus on the Milky Way, by using a detailed thin-disk chemical evolution model (matching local observables, which are weakly affected by radial migration) and a simulation in the cosmological context, with dynamical properties close to those of our Galaxy. We examine in detail the interplay between in situ chemical enrichment and radial migration and their impact on key observables in the solar neighborhood, e.g., the age-metallicity-velocity relation, the metallicity distribution, and gradients in the radial and vertical directions. We show that, due to radial migration from mergers at high redshift and the central bar at later times, a sizable fraction of old metal-poor high-[$\alpha$/Fe] stars can reach the solar vicinity. This naturally accounts for a number of recent observations related to both the thin and thick disks, despite the fact that we use thin-disk chemistry only. Although significant radial mixing is present, the slope in the age-metallicity relation is only weakly affected, with a scatter compatible with recent observational work. While we find a smooth density distribution in the [O/Fe]-[Fe/H] plane, we can recover the observed discontinuity by selecting particles according to kinematic criteria used in high-resolution samples to define the thin and thick disks. We outline a new method for estimating the birth place of the Sun and predict that the most likely radius lies in the range $4.4<r<7.7$~kpc (for a current location at $r=8$~kpc). A new, unifying model for the Milky Way thick disk is offered, where both mergers and radial migration play a role at different stages of the disk evolution. We show that in the absence of early-on massive mergers the vertical velocity dispersion of the oldest stars is underestimated by a factor of $\sim2$ compared with observations. We can, therefore, argue that the Milky Way thick disk is unlikely to have been formed through a quiescent disk evolution. An observational test involving both chemical and kinematic information must be devised to ascertain this possibility.
}

\titlerunning{Chemodynamical evolution of the Milky Way disk I}
\authorrunning{I. Minchev et al.}

\maketitle

\section{Introduction}
\label{sec:intro}

Crucial information regarding the dominant mechanisms responsible for the formation of the Milky Way (MW) disk and other Galactic components is encoded in the chemical and kinematic properties of its stars. This conviction has led to unprecedented observational efforts in the past decades, aimed at mapping the chemistry and kinematics of a large number of stars in the MW. Until the end of 2003 most of the information was confined to small local samples for which high-resolution spectroscopic data were obtained (e.g., \citealt{fuhrmann98} within 25~pc and \citealt{bensby03} within 100~pc). In 2004, the Geneva Copenhagen Survey \cite[CGS][]{nordstrom04} obtained the first large spectro-photometric sample of around 16 000 stars as part of a Hipparcos follow-up campaign (hence, also confined to $\sim$100~pc from the Sun). More recently, optical spectroscopic low-resolution surveys, such as SEGUE \citep{yanny09} and RAVE \citep{steinmetz06}, have extended the studied volume to distances of a few kpc from the Sun (with the majority of stars in the distance range 0.5-3~kpc), and increased the numbers of stars with chemo-kinematical information by more than an order of magnitude ($>2\times10^5$ spectra for SEGUE and $>5\times10^5$ spectra for RAVE, see \citealt{steinmetz12}). This effort will be soon complemented by high-resolution spectroscopic surveys both in the optical -- HERMES \citep{freeman10} and in the near-infrared -- APOGEE \citep{allende08,majewski10}. APOGEE aims at measuring chemo-kinematic properties of around $10^5$ stars close to the Galactic plane, thus complementing SEGUE and RAVE (which exclude most stars below $\sim200$~pc), whereas HERMES aims at obtaining chemical information for around $10^6$ stars. In the near future, 4MOST \citep{dejong12}, probably the most ambitious project, aims at sampling even larger volumes by obtaining chemo-kinematic properties of many millions of stars, taking full advantage of the Gaia astrometric results. The common aim of the huge observational campaigns briefly summarized above is to constrain the MW assembly history -- one of the main goals of the newly emerged field, Galactic Archaeology.

The underlying principle of Galactic Archaeology is that the chemical elements synthesized inside stars, and later ejected back into the interstellar medium (ISM), are incorporated into new generations of stars. As different elements are released into the ISM by stars of different masses and, therefore, on different timescales, stellar abundance ratios provide a cosmic clock, capable of eliciting the past history of star formation and gas accretion of a galaxy\footnote{In most cases, the stellar surface abundances reflect the composition of the interstellar medium at the time of their birth; this is the reason why stars can be seen as fossil records of the Galaxy evolution.}. One of the most widely used ``chemical-clocks" is the [$\alpha$/Fe] ratio\footnote{Here we use the notation in brackets to indicate abundances relative to the Sun, i.e., [X/Y] $= \log$(X/Y)$ - \log$(X/Y)$_{\odot}$.}. 

\subsection{Galactic Archeology and radial migration}

The power of Galactic Archaeology has been threatened both by observational and theoretical results, showing that stars most probably move away from their birthplaces, i.e, migrate radially. Observational signatures of this radial migration (or mixing) have been reported in the literature since the 1970s, with the pioneering works by \cite{grenon72, grenon89}. Grenon identified an old population of \emph {super-metal-rich stars} (hereafter SMR), that are currently located in the solar vicinity, but have kinematics and abundance properties indicative of an origin in the inner Galactic disk (see also \citealt{castro97} and \citealt{trevisan11}). These results were extended by \cite{haywood08}, who showed by re-analyzing the GCS data, that the low- and high-metallicity tails of the thin disk are populated by objects whose orbital properties suggest an origin in the outer and inner Galactic disk, respectively. In particular, the so-called SMR stars show metallicities that exceed the present-day ISM and those of young stars in the solar vicinity. As discussed by \cite{chiappini03} (see also Table~5 by \citealt{asplund09}), the metallicity in the solar vicinity is not expected to have increased much since the Sun's formation, i.e., in the last $\sim$4~Gyr, because of the rather inefficient star formation rate (SFR) at the solar radius during this period, combined with continuous gas infall into the disk. Hence, as summarized in \cite{chiappini09}, pure chemical evolution models for the MW thin disk cannot explain stars more metal rich than $\sim$0.2~dex, and radial migration has to be invoked.

N-body simulations have also long shown that radial migration is unavoidable. \cite{raboud98} studied numerical simulations aimed at explaining the results reported by \cite{grenon89} of a mean positive U-motion (where U is the Galactocentric radial velocity component of stars), which the authors interpreted as metal-rich stars from the inner Galaxy, wandering in the solar neighborhood. However, \cite{raboud98} interpreted their findings as stars on hot bar orbits, not recognizing that permanent changes to the stellar angular momenta are possible. It was not until the work by \cite{sellwood02} that radial migration was established as an important process affecting the entire disk, where stars shift guiding radii due to interaction with transient spiral structure. Modern high-resolution simulations (e.g., \citealt{roskar08a, roskar12}) have confirmed this finding, but left the role of the Galactic bar unexplored. 

A different radial migration mechanism was proposed by \cite{mf10} and \cite{minchev11a}, resulting from the nonlinear coupling between the bar and spiral waves, or spirals of different multiplicity \citep{mq06, minchev12a}. These works, along with studies of diffusion coefficients in barred disks \citep{brunetti11,shevchenko11}, predict a variation in migration efficiency with time and disk radius, establishing that the dynamical influence of the bar plays an integral part in the MW disk modeling. Recently, \cite{comparetta12} showed that radial migration can result from short-lived density peaks arising from interference of spiral density waves, even if patterns are long-lived. In addition to internal axisymmetric structure, perturbations caused by minor mergers have also been shown to be effective at mixing the outer disks \citep{quillen09, bird12}, but can also indirectly affect the entire disk by inducing (reinforcing) spiral and bar instabilities. Considering the established presence of a central bar, spiral structure and evidence for merger activity in the MW, it is clear that all of the above mentioned radial migration mechanisms would have an effect on the Galactic disk.

In summary, a number of both observational and theoretical results challenge the power of Galactic Archaeology. Therefore, the only possible way to advance in this field is the development of chemodynamical models tailored to the MW in the cosmological framework. Only then, a meaningful comparison with the large amounts of current and forthcoming observational data (as summarized in the beginning of this section), can be carried out. This is the main goal of the present work, namely, to develop a chemodynamical model for the MW, to be able to quantify the importance of radial mixing throughout the evolution of our Galaxy. 

\subsection{Difficulties with fully self-consistent simulations}
\label{sec:cos}

Producing disc-dominated galaxies has traditionally been challenging for cosmological models. In early simulations, extreme angular momentum loss during mergers gave birth to galaxies with overly concentrated mass distributions and massive bulges \citep[e.g.,][]{navarro91,navarro94,abadi03}. Although an increase in resolution and better modeling of star formation and feedback have allowed recent simulations to produce MW-mass galaxies with reduced bulge fractions \citep{agertz11, guedes11, martig12}, none of these simulations include chemical evolution. Galaxy formation simulations including some treatment of chemical evolution have been performed by a number groups \citep{raiteri96, mosconi01, lia02, kawata03, kobayashi04, scannapieco05, martinez08, oppenheimer08, wiersma09, few12}. However, although the results are encouraging and globally observed trends seem to be reproduced, such as the mass-metallicity relation \citep[e.g., ][]{kobayashi07} or the metallicity trends between the different Galactic components \citep[e.g., ][]{tissera12}, it is still a challenge for these simulations to match the properties of the MW (e.g., the typical metallicities of the different components -- Ê\citealt{tissera12}). Additionally, the fraction of low-metallicity stars are often overestimated \citep{kobayashi11, calura12}, and reproducing the position of thin- and thick-disk stars in the [O/Fe]-[Fe/H] plane has proved challenging \citep{brook12}. While these problems might be due to unresolved metal mixing \citep{wiersma09}, it is also worth noting that none of the above-mentioned simulations reproduces simultaneously the mass, the morphology, and the star formation history (SFH) of the MW.

This situation has led us to seek a novel approach to address this complex problem. We will show that this approach works encouragingly well, explaining not only current observations, but also leading to a more clear picture regarding the nature of the MW thick disk. 

\subsection{Thick-disk formation scenarios}

The large uncertainties in important observational constraints in the MW, such as the age-velocity-metallicity relation, the abundance gradients and their evolution, together with the inherent complexity of the topic of Galaxy assembly, have led to different scenarios to be proposed for the formation of the thick disk. 

One possibility is that thick disks were born thick at high redshift from the internal gravitational instabilities in gas-rich, turbulent, clumpy disks \citep{bournaud09, forbes12} or in the turbulent phase associated with numerous gas-rich mergers \citep{brook04,brook05}. They could also have been created through accretion of galaxy satellites \citep{meza05,abadi03}, where thick-disk stars then have an extragalactic origin. 

Another possibility is that thick disks are created through the heating of preexisting thin disks with the help of mergers \citep{quinn93, villalobos08, dimatteo11}, whose rate decreases with decreasing redshift. Evidence for satellite-disk encounters can be found in structure in the phase-space of MW disk stars (e.g., \citealt{minchev09, gomez13, gomez12c}), which can last for as long as $\sim$4~Gyr \citep{gomez12b}. 

Finally, a recently proposed mechanism for the formation of thick disks is radial migration, which we discuss next. 

\subsection{Radial migration and thick disks}
\label{sec:mig}

In the past several years there has been a growing conviction that radial migration, driven by transient spirals, can be responsible for the formation of thick disks by bringing out high-velocity-dispersion stars from the inner disk and the bulge (no need for mergers). This scenario was used, for example, in the analytical model of \cite{schonrich09a} (SB09a) and \cite{schonrich09b} (SB09b), where the authors managed to explain the MW thin- and thick-disk characteristics without the need of mergers or any discrete heating processes. Similarly, the increase of disk thickness with time found in the simulation by \cite{roskar08a} has been attributed to migration in the works by \cite{sales09} and \cite{loebman11}. 

However, how exactly radial migration affects disk thickening in dynamical models had not been demonstrated (but only assumed) until the recent work by \cite{minchev11b} and \cite{minchev12b}. The latter authors showed unambiguously that radial migration driven by secular (or internal) evolution has only a minor effect on disk thickening, mostly beyond three disk scale-lengths, which results in a flared disc\footnote{\cite{minchev12b} showed that migrators {\it contract} (vertically cool) the disk inside the bar's CR, further contributing to disk flaring. This is related to the predominance of inward migrators at such small radii with a mean action smaller than that of the local population.}. This is due to the conservation of vertical action of migrators, as opposed to the incorrect assumption of vertical energy conservation. While not building up a {\it kinematically} thick disk (since the effect of outward and inward migrators mostly cancels out), \cite{minchev12b} noted that radial migration in isolated galaxies does contribute to a {\it chemically} thick disk in the sense that outward migrators are preferentially deposited at higher distances above the galactic plane, with the converse effect for inward migrators. It can be expected that migration {\it can} thicken the disk, provided that stars were ``preheated", e.g., either by mergers or by being born hot\footnote{Note that migration efficiency drops with velocity dispersion \citep{solway12}.} (both of these are expected at high redshift). 

Once again, to advance our understanding of the MW disk formation and evolution, we need to make quantitative estimates of the importance of radial mixing, guided by observational constraints. This implies the need for chemodynamical models making predictions specifically for the solar vicinity, where most of the current observational constraints are found. In Sec.~\ref{sec:sim} we describe the disk simulation we adopt in the present work. Sec.~\ref{sec:chem} we describe our input chemistry. Sec.~\ref{sec:tech} is devoted to our new approach. Our results are shown in Sec.~\ref{sec:res}, while a new explanation for the origin of the thick disk can be found in Sec.~\ref{sec:thick}. Here we concentrate on results for the solar vicinity, while results for the whole disk are shown in paper II of this series. Conclusions are drawn in Sec.~\ref{sec:concl}.

\begin{figure*}
\includegraphics[width=18cm]{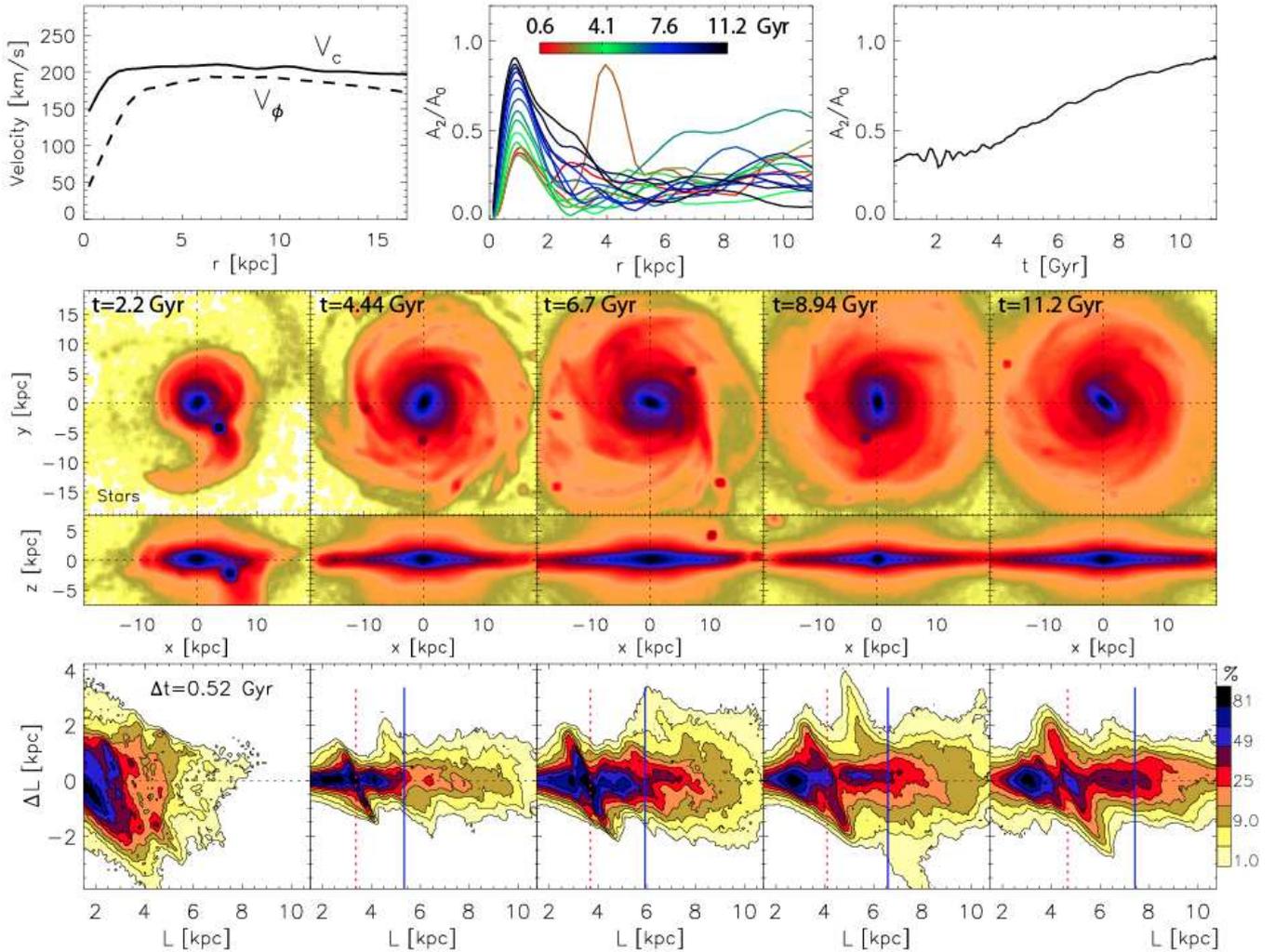}
\caption{
{\bf First row:} The left panel shows the rotational velocity (dashed blue curve) and circular velocity (solid black curve) at the final simulation time. The middle panel presents the $m=2$ Fourier amplitudes, $A_2/A_0$, as a function of radius estimated from the stellar density. Curves of different colors present the time evolution of $A_2/A_0$. To better see the evolution of the bar strength with time, in the right panel we show the amplitude averaged over the bar maximum. {\bf Second row:} Face-on density maps of the stellar component for different times, as indicated. {\bf Third row:} The corresponding edge-on view. Contour spacing is logarithmic. {\bf Fourth row:} Changes in angular momentum, $\Delta L$, as a function of radius, estimated in a time window of 0.52~Gyr, centered on the times of the snapshots shown above. Both axes are divided by the circular velocity, thus units are kpc (galactic radius). Strong variations are seen with cosmic time due to satellite perturbations and increase in bar strength. 
}
\label{fig:xy}      
\end{figure*}

\section{A late-type disk galaxy simulation in the cosmological context}
\label{sec:sim}

To properly model the MW, it is crucial to be consistent with some observational constraints at redshift $z=0$, for example, a flat rotation curve (e.g., \citealt{merrifield92}), a small bulge (e.g., \citealt{binney97}), a central bar of an intermediate size (e.g., \citealt{babusiaux05}), gas-to-total-disk-mass ratio of $\sim0.14$ at the solar vicinity (e.g., \citealt{chiappini01}, and references therein), and local disk velocity dispersions close to the observed ones (e.g., \citealt{lee11}).

It is clear that cosmological simulations would be the natural framework for a state-of-the-art chemodynamical study of the MW. Unfortunately, as described in Sec.~\ref{sec:cos}, a number of star formation and chemical enrichment problems still exist in fully self-consistent simulations. We therefore resort to (in our view) the next best thing -- a high-resolution simulation in the cosmological context coupled with a pure chemical evolution model (Sec.~\ref{sec:chem}), as described in detail in Sec.~\ref{sec:tech} below.

The simulation used in this work is part of a suite of numerical experiments first presented by \cite{martig12}, where the authors studied the evolution of 33 simulated galaxies from $z=5$ to $z=0$ using the zoom-in technique described in \cite{martig09}. This technique consists of extracting merger and accretion histories (and geometry) for a given halo in a $\Lambda$-CDM cosmological simulation, and then re-simulating these histories at much higher resolution (150~pc spatial, and 10$^{4-5}$~M$_{\odot}$ mass resolution). We refer to \cite{martig12} for more information on the simulation method. 

The galaxy we have chosen has a number of properties consistent with the MW, including a central bar, as we describe below. We would like to explore the disk evolution for a time period of about 11~Gyr, which is close to the age of the oldest disk stars in the MW. However, none of the above galaxies start forming disks earlier than $\sim9-10$~Gyr before $z=0$. To remedy this, we extended our simulation by integrating for two additional Gyr, allowing for 11.2~Gyr of self-consistent evolution.

Originally, our simulated galaxy has a rotational velocity at the solar radius of 210 km/s and a scale-length of $\sim5$~kpc. To match the MW in terms of dynamics, at the end of the simulation we downscaled the disk radius by a factor of 1.67 and adjusted the rotational velocity at the solar radius to be 220 km/s, which affects the mass of each particle according to the relation $GM\sim v^2r$, where $G$ is the gravitational constant. This places the bar's corotation resonance (CR) and 2:1 outer Lindblad resonance (OLR) at $\sim4.7$ and $\sim7.5$~kpc, respectively, consistent with a number of studies\footnote{Variations in our results with different rescaling are discussed in appendix A.} (e.g., \citealt{dehnen00}, \citealt{mnq07, minchev10}, see \citealt{gerhard11} for a review). At the same time the disk scale-length, measured from particles of all ages in the range $3<r<15$~kpc, becomes $\sim3$~kpc, in close agreement with observations (e.g., \citealt{gerhard01}). After this change{\footnote{Note that we rescaled the disk radius to reflect the position of the bar's resonances at the final time.}, our simulated disk satisfies the criteria outlined at the beginning of this section, required for any dynamical study of the MW, as follows:

(i) it has an approximately flat rotation curve with a circular velocity $V_c\sim220$~km/s at 8~kpc, shown in the first row, left panel of Fig.~\ref{fig:xy}, where we have corrected for asymmetric drift as described in, e.g., \cite{bt08}. 

(ii) the bulge is relatively small, with a bulge-to-total ratio of $\sim$1/5 (as measured with GALFIT -- \citealt{peng02} -- on a mock i-band image, see bottom panel of Fig.~27 by \citealt{martig12}). 

(iii) it contains an intermediate-size bar at the final simulation time, which develops early on and grows in strength during the disk evolution (see middle and right panels of Fig.~\ref{fig:xy} and discussion below).

(iv) the disk grows self-consistently as the result of cosmological gas accretion from filaments and (a small number of) early-on gas-rich mergers, as well as merger debris, with a last significant merger concluding $\sim9-8$~Gyr ago. 

(v) the disk gas-to-total-mass ratio at the final time is $\sim$0.12, consistent with the estimate of $\sim0.14$ for the solar vicinity (Fig.~\ref{fig:sfh_sn}, top right panel). 

(vi) the radial and vertical velocity dispersions at $r\approx8$~kpc are $\sim45$ and $\sim20$~km/s (averaged over all ages), in very good agreement with observations (see Sec.~\ref{sec:sig} and Fig.~\ref{fig:sig}).

We defined $t=0$ as the time after the central bulge (spheroidal component) has formed. Only about 3\% of these stars contribute to the stellar density at the final simulation time in the region $7<r<9$~kpc, $|z|<3$~kpc, which we investigate in this paper. Therefore, for this work we ignored these old stars and focused on the particles forming mostly from gas infall at later times.  

In the first row, middle panel of Fig.~\ref{fig:xy}, we plot the Fourier amplitude, $A_m/A_0$, of the density of stars as a function of radius, where $A_0$ is the axisymmetric component and $m$ is the multiplicity of the pattern; here we only show the $m=2$ component, $A_2/A_0$. Different colors indicate the evolution of the $A_2$ radial profile in the time period specified by the color bar. The bar is seen to extend to $\sim3-4$~kpc, where deviations from zero beyond that radius are due to spiral structure. The brown curve reaching $A_2/A_0\sim0.9$ at $r\approx4.5$~kpc results from a merger-induced two-armed spiral. To better see the evolution of the bar strength with time, in the right panel we show the amplitude averaged over 1~kpc at the bar maximum. 

The second and third rows of Fig.~\ref{fig:xy} show face-on and edge-on stellar density contours at different times of the disk evolution, as indicated in each panel. The redistribution of stellar angular momentum, $L$, in the disk as a function of time is shown in the fourth row, where $\Delta L$ is the change in the specific angular momentum as a function of radius estimated in a time period $\Delta t=520$~Myr, centered on the time of each snapshot above. Both axes are divided by the rotational velocity at each radius, therefore $L$ is approximately equal to the initial radius (at the beginning of each time interval) and $\Delta L$ gives the distance by which guiding radii change during the time interval $\Delta t$. The dotted-red and solid-blue vertical lines indicate the positions of the bar's CR and OLR. Note that due to the bar's slowing down, these resonances are shifted outward in the disk with time.

After the initial bulge formation, the largest merger has a 1:5 stellar-mass ratio and an initially prograde orbit, plunging through the center later and dissolving in $\sim1$~Gyr (in the time period $1.5\lesssim t\lesssim2.5$~Gyr, first column of Fig.~\ref{fig:xy}). Due to its in-plane orbit (inclination $\lesssim45^\circ$), this merger event results in accelerated disk growth by triggering a strong spiral structure in the gas-dominated early disk, in addition to its tidal perturbation \citep{quillen09}. One can see the drastic effect on the changes in angular momentum, $\Delta L$, in the fourth row, right panel of Fig.~\ref{fig:xy}. A number of less violent events are present at that early epoch, with their frequency decreasing with time. The effect of small satellites, occasionally penetrating the disk at later times, can be seen in the third and fourth columns at $L\approx r\gtrsim7$ and $L\approx r\approx6$~kpc, respectively. 

We note a strong variation of $\Delta L$ with cosmic time, where mergers dominate at earlier times (high $z$) and internal evolution takes over at $t=5-6$~Gyr (corresponding to a look-back-time of $\sim$6-7~Gyr, or $z\sim1$). The latter is related to an increase with time in the bar's length and major-to-minor-axes ratio as seen in the face-on plots, indicating the strengthening of this structure. Examining the top-right panel of Fig.~\ref{fig:xy}, we find a continuous increase in the bar's $m=2$ Fourier amplitude with time, where the strongest growth occurs between $t$=4 and 8~Gyr. The effect of the bar can be found in the changes in angular momentum, $\Delta L$, as the feature of negative slope, centered on the CR (dotted-red vertical line), shifting from $L\approx3.4$ to $L\approx4.7$~kpc. Due to the increase in the bar's amplitude, the changes in stellar guiding radii (vertical axis values) induced by its presence in the CR-region double in the time period  $4.44<t<11.2$~Gyr (bottom row of Fig.~\ref{fig:xy}). Until recently, bars were not considered effective at disk mixing once they were formed, because of their long-lived nature. We emphasize the importance of the bar in its persistent mixing of the inner disk {\it throughout the galactic evolution} (see \citealt{minchev12a} and discussion therein).

\section{Chemistry}
\label{sec:chem}

Our new approach is based on a detailed chemical evolution model for the \emph{thin disk only} (see appendix B). The idea behind this is to test, after radial mixing is taken into account, whether we can explain the observations of both thin and thick disks without the need of invoking a discrete thick-disk component \cite[different from what was suggested in][]{chiappini97,chiappini09}.

In the present model, the thin disk forms via the slow accretion of extragalactic metal-poor gas (assumed to be of primordial composition). We assumed that the gas accumulates faster in the inner than in the outer regions, similar to what happens in the simulation discussed in Sec.~\ref{sec:sim} (the exact functional form used for gas infall as a function of both time and radius is shown in appendix B). The present model does not account for gas outflows or gas radial flows. Our code follows in detail a large number of chemical elements by properly taking into account the lifetime of stars of different masses (i.e., we did not use the instantaneous recycling approximation). We here concentrate only on iron and oxygen (see appendix B for a detailed description of the adopted stellar yields, as well as how we treated the supernova (SN) Type Ia enrichment in our model)\footnote{We often refer to the [$\alpha$/Fe] ratios in the text, where in our case oxygen is the chosen $\alpha$ element.}, while other chemical elements are discussed in the forthcoming papers of this series. In all plots in this paper the results are normalized using the solar abundances by \cite{asplund05}.

\begin{figure}
\includegraphics[width=8.5cm]{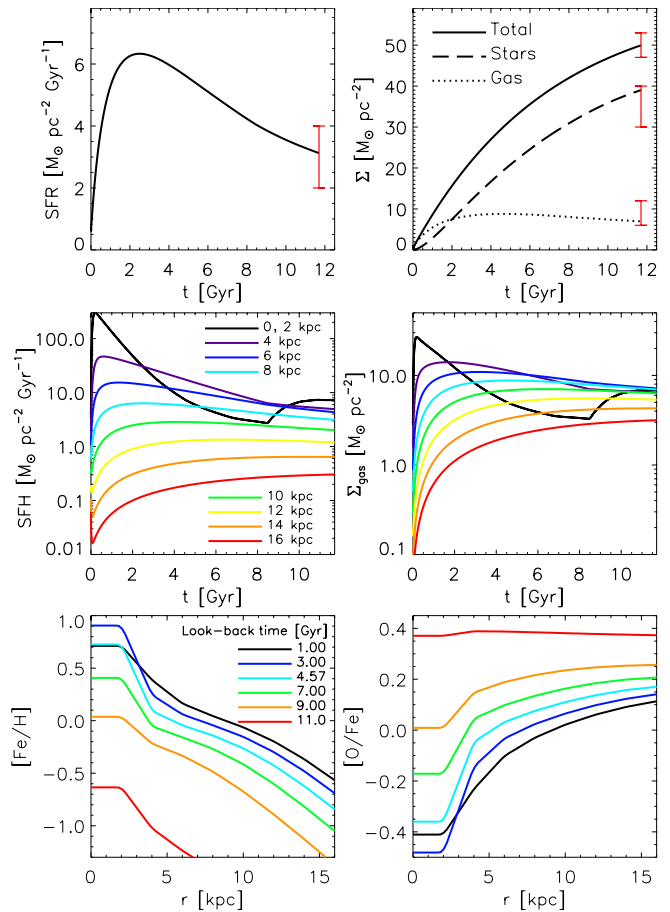}
\caption{
Properties of our detailed thin-disk chemical evolution model: 
{\bf Top row, left:} SFR in the solar neighborhood resulting from our model. {\bf Top row, right:} The total (solid), stellar (dashed), and gas (dotted) density time evolution in the solar neighborhood. The error bars are observational constraints (see Table B.1 for references to the data).
{\bf Middle row, left:} SFR as a function of time for different galactic radii as color-coded. 
{\bf Middle row, right:} Total gas density (remaining after star formation + recycled) as a function of time for different galactic radii as color-coded on the left. 
{\bf Bottom row, left:} [Fe/H] gradients. Different curves and colors correspond to different look-back times as indicated.
{\bf Bottom row, right:} [O/Fe] gradients for different look-back times as on the left.
}
\label{fig:sfh_sn}      
\end{figure}

The main observational constraints of MW chemical evolution models (see Table~B.1 in appendix B) are (i) the solar and present-day abundances of more than 30 elements, (ii) the present SFR (Fig.~\ref{fig:sfh_sn}, upper left panel), (iii) the current stellar, gas, and total mass densities in the solar vicinity (Fig.~\ref{fig:sfh_sn}, upper right panel), (iv) the present-day supernovae (SNe) rates of Type II and Ia, and (v) the metallicity distribution of G-dwarf stars (to be discussed in Sec.~\ref{sec:res}). The SFH in the solar vicinity (Fig.~\ref{fig:sfh_sn}, upper left panel) is obtained by assuming a star formation law dependency on the gas density (Kennicut's law) and an exponentially decreasing infall with time, with an infall timescale $\tau$(R$_{g,\odot})=7$~Gyr, where R$_{g,\odot}=8$~kpc is the Sun's galactocentric distance (see appendix B for more details). From the chemical evolution constraints listed above, the one most affected by radial migration is the metallicity distribution\footnote{Note that the age-metallicity relation, also affected by radial migration, was never used as a constraint in our chemical evolution models, because of the large scatter shown in the data.}. 

Outside the solar vicinity chemical evolution models are mainly constrained by the present-day abundance gradients in the thin disk and are, thus, more uncertain than the model for the solar vicinity (see discussion in \citealt{chiappini01}). \cite{chiappini01} emphasized that the radial abundance gradients predicted along the thin disk, as well as their time evolution, are strongly dependent on pre-enrichment and star formation threshold assumptions. Previous chemical evolution models in the literature considering the thin disk as an independent component could not anticipate such effects. Here we keep it simple by assuming neither threshold nor pre-enrichment; as a consequence, our models predict flattening of the gradients with time, similar to other pure thin-disk chemical evolution models in the literature (e.g., \citealt{hou01}). On the observational side, the question of whether the abundance gradients steepen or flatten with time is still not settled.

The chemical gradients we obtained are the result of the interplay between the infall and star formation rates at different radii. Here we used the same $\tau(r)$ expression as in \cite{chiappini01} down to $r=2$~kpc (details can be found in appendix B). Chemical evolution models are often not computed for the innermost 4~kpc to avoid dealing with the complete lack of observational constraints and with a region where different Galactic components co-exist (bulge, bar, and inner disc). Because here our main focus is the disk, we would like to assign chemistry down to $r=0$ and made the conservative hypothesis that the chemistry computed for $r=2$~kpc applies to all particles with $r<2$~kpc, as well.

The resulting SFRs at different galactocentric radii are shown in the middle-left panel of Fig.~\ref{fig:sfh_sn}, while the gas density (remaining after stars formation + recycled) is plotted on the right. The resulting gas metallicity and [O/Fe] gradients for different look-back times can be found in the bottom row of the same figure.

Currently, there are still large uncertainties also in the shape of present-day abundance gradients. From a recent compilation of the best data available in the literature, \cite{stasinska12} concluded that most claims of steepening or flattening of the \emph{current} abundance gradients toward the center or the outskirts of the MW are premature. A flattening of the disk planetary nebulae abundance gradient towards the centre of the Galaxy is, however, indicated by the comparison of the works by \cite{chiappini09} and \cite{henry10}. Whether this applies to other abundance gradient tracers is, as of yet, not established. Indeed, \cite{pedicelli09} found that Cepheids located in the inner disk appear systematically more metal-rich than the mean metallicity gradient. A flattening of the MW abundance gradients in the outer regions of the MW is also suggested by the Cepheid data (see discussion in \citealt{cescutti07}). It is worth noticing that in some external galaxies a linear gradient was found in the inner regions, while a flattening is observed toward large distances from the center (e.g \citealt{bresolin12} and references therein).

With our current SFH assumptions, the predicted present-time iron gradient, shown by the black curve in the bottom-left panel of Fig.~\ref{fig:sfh_sn}, amounts to $\approx-0.061$ or $\approx-0.057$~dex/kpc, depending on the radial range we used for fitting (5-12 or 6-11~kpc, respectively - see Fig. B.2 for the predicted oxygen abundance gradient compared with that of Cepheids).  As we show below, slightly flatter values are obtained ($\approx-0.059$ or $\approx-0.058$~dex/kpc) after radial migration is taken into account (Sec.~\ref{sec:res}). It is beyond the scope of the present work to check for the impact of other assumptions on the abundance gradients, but this will certainly be the focus of a future paper of this series. If the initial gradients turn out to be flatter than those adopted here, the effects of radial migration in some observables (as the age-metallicity relation) will, of course, be weaker than those described in Sec.~\ref{sec:grad}.

\section{New approach for combining chemistry and dynamics}
\label{sec:tech}

Despite the recent advances in the general field of galaxy formation and evolution, there are currently no self-consistent simulations that have the level of chemical implementation required for making detailed predictions for the number of ongoing and planned MW observational campaigns. Even in high-resolution N-body experiments, one particle represents $\sim10^{4-5}$ solar masses. Hence, a number of approximations are necessary to compute the chemical enrichment self-consistently inside a simulation, leading to the several problems briefly discussed in the introduction. Here, instead, we assumed that each particle is one star\footnote{Dynamically, this is a good assumption, since the stellar dynamics is collisionless.} and implemented the exact SFH and chemical enrichment from our chemical model (Sec.~\ref{sec:chem}) into our simulated galactic disk (Sec.~\ref{sec:sim}). This can be thought of as inserting the dynamics of our simulation into the chemical model, not the chemistry into the simulation. 

We started by dividing the disk into 300-pc radial bins in the range $0<r<16$~kpc. At each time output we randomly selected newly born stars by matching the SFH corresponding to our chemical evolution model at each radial bin. Since our goal is to satisfy the SFH expected for the chemical model (similar, but not identical to that of the simulation), at some radial bins and times we may not have enough newly born stars in the simulation. In that case we randomly selected stars currently on the coldest orbits (i.e., with kinematics similar to those of newly born stars). This fraction is about 10\% at the final time. The particles selected at each time output were assigned the corresponding chemistry for that particular radius and time. This was repeated at each time-step of 37.5~Myr. For the current work we extracted about 40\% of the $\approx2.5x10^6$ stellar particles comprising our dynamical disk at the final time in the range $|z|<4$~kpc, $r<16$~kpc. We verified that decreasing or increasing this extracted sample by a factor of two had negligible effects on the results we present in this work.

In the above described manner, we followed the self-consistent disk evolution for 11.2~Gyr, selecting a tracer population with known SFH and chemistry enrichment. This bypasses all the problems encountered by previous chemodynamical models based on N-body simulations. It also offers a way to easily test the impact of different chemical (or dynamical, see appendix A) prescriptions on the chemodynamical results. 

From a chemical point of view, there are two main simplifications in our approach. Firstly, we neglected radial gas flows, Galactic fountains, and SN-driven winds, which may be resulting in flatter abundance gradients than we currently find (see also \citealt{spitoni11}). This issue will be addressed in paper II. Second, we assumed that stars do not contribute to the chemical evolution beyond the zone where they are born, but either contribute only to the chemical enrichment within 2~kpc from their birth place, or never die. This assumption is valid for most of the stars because (i) the massive stars die essentially where they were born due to their short lifetimes and (ii) low-mass stars live longer than the age of the Galaxy (never die). We do not expect this simplification to affect our results by more than 10\% for chemical elements made in low- and intermediate-mass stars and even less for those coming from massive stars, as is the case for oxygen.

From a dynamical point of view, the main simplifications are (i) the resampling of the simulation SFH according to the chemical evolution model and (ii) the difference between the gas-to-total-mass ratio expected from the chemical model and attained by the simulation. Although the disks in both the chemical and dynamical models grow inside-out, there are some offsets at particular times. These differences in the SFHs are unavoidable, since the chemical and dynamical models are not tuned to reproduce the same star formation, although they are quite similar for most of the evolution (see also appendix A). 

As already mentioned in Sec.~\ref{sec:sim}, the disk gas-to-total-mass ratio at the final time is $\sim$0.12, consistent with the estimate of $\sim0.14$ for the solar vicinity (Fig.~\ref{fig:sfh_sn}, top-right panel). However, at earlier times the gas fraction at the solar location, for example, can be overestimated by a factor of about 1.5-2 with respect to the chemical model. One effect of a larger gas fraction can be a decreased susceptibility to merger perturbations (e.g., \citealt{moster10}). It may also give rise to a more unstable disk, while suppressing the bar instability (e.g., \citealt{bournaud02}). Indeed, at earlier times we do find a weaker bar that appears more like a lens, which strengthens later, showing an increase in both its length and major-to-minor-axes ratio (Fig.~\ref{fig:xy}). On the other hand, unreasonable disk instabilities due to a large gas fraction in the outer disk are not seen in our simulation. In fact, the spiral structure is about 15\% of the background density and does not vary considerably, except when reinforced by mergers at earlier epochs. This amplitude is similar to expectations for the MW (e.g., \citealt{drimmel01,siebert12}). 

While the difference between the assumed (chemical model) and actual (dynamical model) stellar and gas densities can introduce some inconsistencies in the resulting dynamics, this would generally have the tendency of bringing more stars from the inner disk out, due to a larger bar expected at earlier times. As we show below, a larger fraction of old stars coming from the bar's CR region to the solar vicinity would only strengthen our results.  

We note that the rescaling of the disk at the end of the simulation would result in deviations from the original cosmological prescription, because satellite orbits would appear more central after downscaling by the factor of 1.67 (see Sec.~\ref{sec:sim}). It is hard to predict how much the overall disk dynamics would be affected by this, but it should be kept in mind that mergers and gas inflows become increasingly unimportant with decreasing redshift.

Overall, we do not anticipate the simplifications of our approach to significantly affect any of our results (see appendix A). On the other hand, what we gain with this technique is a new tool to study the chemodynamics of our Galaxy, which is complementary to fully self-consistent models, and where the overall complex problem of Galaxy assembly and evolution can be understood by pieces (i.e., same chemistry applied to different simulations and different chemistry applied to the same simulation). We anticipate that this new approach will also be useful to gain insights that can later be used in fully self-consistent simulations.

\begin{figure*}
\includegraphics[width=18cm]{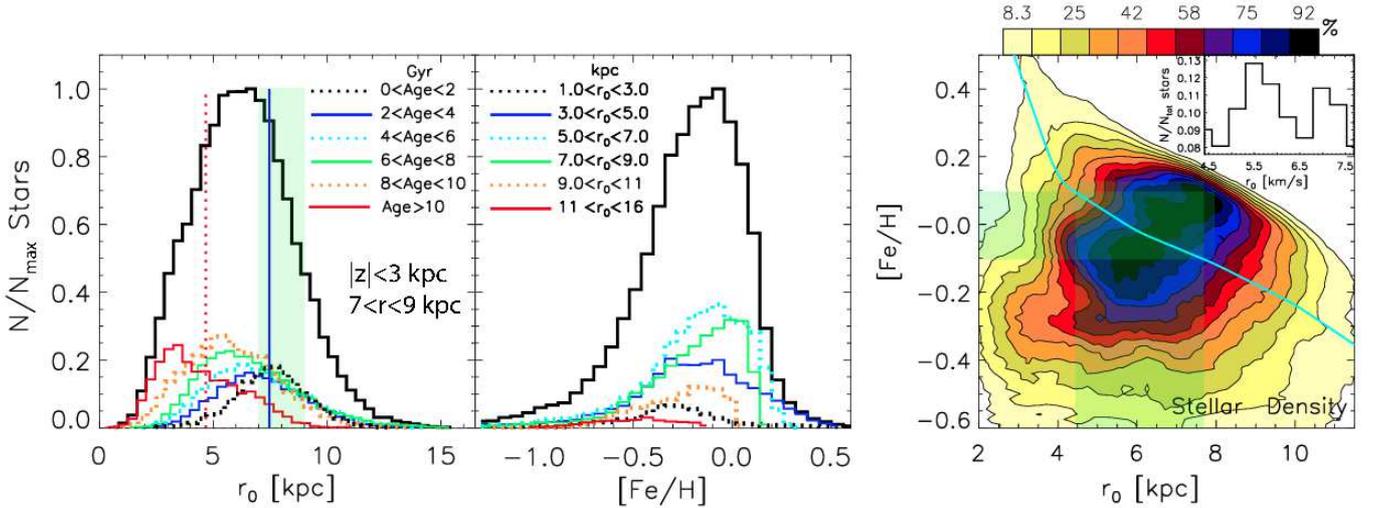}
\caption{
{\bf Left:} Birth radii of stars ending up in the ``solar" radius (green shaded strip) at the final simulation time. The solid black curve plots the total $r_0$-distribution, while the color-coded curves show the distributions of stars in six different age groups, as indicated. The dotted-red and solid-blue vertical lines indicate the positions of the bar's CR and OLR at the final simulation time. A large fraction of old stars comes from the inner disk, including from inside the CR. {\bf Middle:} [Fe/H] distributions for stars ending up in the green-shaded strip (left) binned by birth radii in six groups, as indicated. The total distribution is shown by the solid black curve. The importance of the bar's CR is seen in the large fraction of stars with $3<r_0<5$~kpc (blue line). {\bf Right:} Density contours of the $r_0$-[Fe/H] plane for local stars. The cyan curve shows our model solar-age metallicity gradient. Taking an error of $\pm1$~dex in [Fe/H], we find a possible Sun birth radius of $4.6<r_0<7.6$~kpc (where the horizontal green, transparent strip meets the cyan line). The imbedded histogram shows the density of stars in the likely $r_0$-range, indicating a decline in the probability for $r_0\lesssim5.5$~kpc.  
}
\label{fig:r0}      
\end{figure*}

\section{The resulting chemodynamics}
\label{sec:res}

After fusing the chemical and dynamical models as outlined above, we are now in a position to (i) look for deviations from the predictions of the purely chemical evolution model and (ii) investigate the causes for these differences. 

To first illustrate the migration efficiency in our simulation, in the left panel of Fig.~\ref{fig:r0} we show the birth radii of stars ending up in a solar neighborhood-like location ($7<r<9$~kpc, green shaded strip, and $|z|<3$~kpc) after 11.2~Gyr of evolution. The bar's CR and OLR (estimated at the final time) are shown by the dotted-red and solid-blue vertical lines. The solid black line plots the total population, which peaks close to $r_0=6$~kpc due to radial migration. An overdensity is seen just inward of the bar's CR. The entire sample is also divided into six age-groups, shown by the curves of different colors and line-styles. The strongest effect from radial migration is found for the oldest stars (red curve), whose distribution has a maximum at $r\approx3$~kpc, or inside the bar's CR. Note that locally born stars of all ages can be found in the solar neighborhood. A relatively smooth transition of the peak from older to younger groups of stars is observed; this is expected, since even for a constant migration efficiency older stars would be exposed longer to perturbations. While a wide range of birth radii is seen for all age groups, the majority of youngest stars were born at, or close to, the solar neighborhood bin.

\begin{figure*}
\includegraphics[width=18cm]{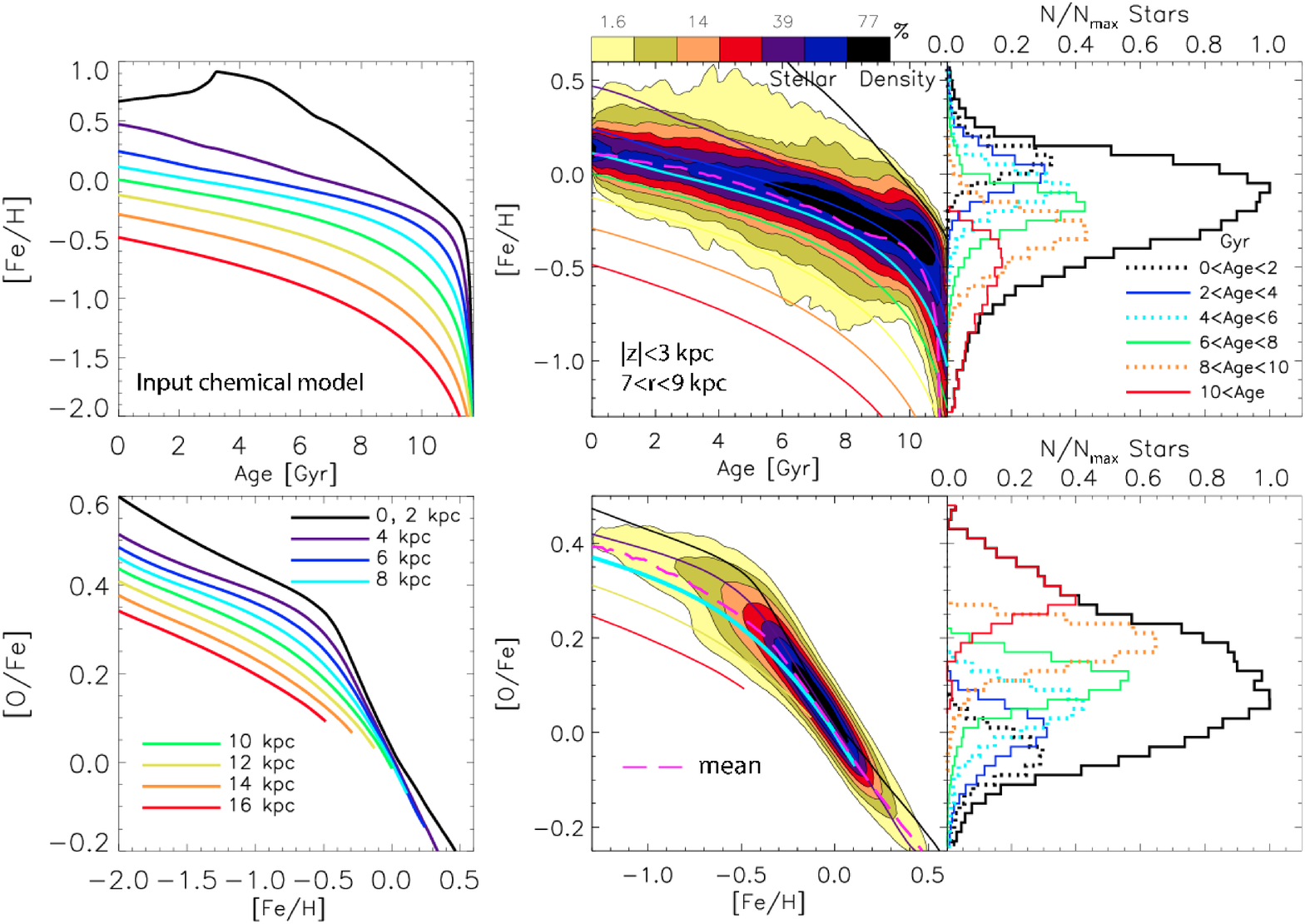}
\caption{
{\bf Top:} The left panel plots age versus [Fe/H] for different radii, resulting from our input chemical model. The middle panel shows stellar density contours of the resulting relation after fusing with dynamics, for the ``solar" radius ($7<r<9$~kpc). The overlaid curves indicate the input chemistry for some of the radii shown in the left panel. The pink dashed curve plots the mean, seen to closely follow the input $r_0=8$~kpc line (cyan). The right panel plots the metallicity distributions for different age bins. {\bf Bottom:} Same as above, but for the [Fe/H]-[O/Fe] relation.
}
\label{fig:chem}      
\end{figure*}

\subsection{The metallicity distribution}

The middle panel of Fig.~\ref{fig:r0} presents the metallicity distribution of stars in our simulated solar neighborhood (green-shaded strip in left panel). In addition to the total distribution (solid black line), six subsets of stars grouped by birth radii are shown by the different colors and line-styles. The locally born sample (green line) peaks slightly beyond [Fe/H]=0, ends abruptly at [Fe/H]$\sim0.15$ and has an extended tail toward lower metallicities. This distribution is the same as the one predicted by our pure-thin-disk chemical evolution model with the SFH shown in the top-left panel of Fig.~\ref{fig:sfh_sn}. Our chemodynamical model, on the other hand, predicts that the majority of migrators comes from the inner region ($3<r_0<5$ and $5<r_0<7$~kpc bins), as expected. A smaller, but sizable, fraction of stars originating {\it outside the solar neighborhood} is also observed (orange and red curves). 

While the low-metallicity end in the total distribution comes from contribution by all initial radii, the tail at higher metallicities extending to [Fe/H]$\sim0.6$ results exclusively from stars with $3<r_0<5$~kpc (note that the $5<r_0<7$~kpc bin contributes up to [Fe/H]$\sim0.25$ only). This is the region just inside the bar's CR, which is where the strongest outward radial migration occurs, as we discussed in Sec.~\ref{sec:sim} and showed in the bottom row of Fig.~\ref{fig:xy}. At smaller radii stars are trapped in the $x_1$ bar orbits (e.g., \citealt{contopoulos80}), making outward angular momentum transfer difficult. Nevertheless, stars originating from $1<r_0<3$~kpc still appear in the local bin, with a fraction similar to that of objects coming from $r_0>11$~kpc. We discuss how the metallicity distribution relates to observations in Sec.~\ref{sec:segue} (Figs.~\ref{fig:mdf} and \ref{fig:mdf1}).

\subsection{The birth place of the Sun}

With the care taken in defining a proper solar radius in a simulation with MW characteristics, combined with MW chemistry, it is tempting to make an estimate for the solar birth radius, which emerges naturally from our chemodynamical model.
 
Already from the left panel of Fig.~\ref{fig:r0} we can see that the age distribution to which the Sun belongs (dotted-cyan curve) suggests a most likely solar birth location at $r\approx6$~kpc and similar probability of it being born in situ or at $r\approx5$~kpc. By including the chemical information we can further improve this estimate. 

The right panel of Fig.~\ref{fig:r0} displays density contours of the $r_0$-[Fe/H] plane for all local stars. The cyan curve shows our input solar-age (4.6~Gyr look-back time) metallicity gradient. Assuming an error of $\pm1$~dex in [Fe/H], we find a possible Sun birth radius in the range $4.4<r_0<7.7$~kpc (where the horizontal green transparent strip meets the cyan curve). 

The imbedded $r_0$-histogram in the right panel of Fig.~\ref{fig:r0} shows the density of stars in the likely $r_0$-range, estimated from stars in a narrow age-bin around the solar value ($4.6\pm0.1$~Gyr, consistent with, e.g., \citealt{bonanno02,christensen09,houdek11}). We find the highest probability to be around 5.6~kpc, followed by 7~kpc. Note that this estimate is dependent on the migration efficiency in our simulation and the adopted chemical evolution model. Our result agrees well with the estimate of $6.6\pm0.9$~kpc by \cite{wielen96}.

\subsection{The age-[Fe/H] and [Fe/H]-[O/Fe] relations}
\label{sec:amr}

We showed in Figs.~\ref{fig:xy} (bottom) and \ref{fig:r0} (left) that radial mixing is significant in the simulation we use here. It is therefore very interesting to find out how much the age-metallicity (AMR) and [Fe/H]-[O/Fe] relations are affected.

To understand our results better, we show in the first row, left panel of Fig.~\ref{fig:chem} the metallicity time variation of our input chemical model for different initial radii, as indicated in the lower panel. The middle top panel shows stellar density contours of the resulting relation after fusing with dynamics for the ``solar" cylinder ($7<r<9$~kpc, $|z|<3$~kpc). Every other curve from the left panel is overlaid on top of the contours, giving an insight into the origin of stars found in this localized region. The excess of stars obove the local curve (cyan) is due to migrators coming from the inner disk, as seen in Fig.~\ref{fig:r0}. Similarly, contours below the local curve result from stars originating in the outer disc\footnote{The deficiency of stars seen at ages~$\gtrsim9$~Gyr ([Fe/H]$\approx-0.8$) in the lowest-density contours is an artifact due to the slightly lower SFR in our simulation at $r>10$~kpc; note that this only affects a negligibly small number of the oldest stars at this large radius.}. 

The pink dashed curve plots the mean metallicity, which is found to follow closely the in-situ born population (cyan curve). Some flattening is observed, mostly for ages $\gtrsim$9~Gyr, but the final distribution is by no means flat. The reason for this minor effect on the local metallicity gradient, despite the strong migration, is the fact that at the Sun's intermediate distance from the Galactic center the change in metallicity arising from stars migrating from the inner regions is mostly compensated for by stars arriving from the outer disk.

The bump seen in the contours and the mean (pink line) at ages~$\sim10$~Gyr and [Fe/H]~$\approx-0.2$ results from the strong merger-related migration at that time (see Sec.~\ref{sec:liu}). 
Identifying such a structure in the observed AMR would be indicative of a strong change in the migration efficiency with time. 

The right top panel of Fig.~\ref{fig:chem} shows the metallicity distributions for different age bins. This is consistent with recent observational results (e.g., \citealt{haywood12} and references therein) in showing that the scatter in metallicity increases toward older ages. We find the following dispersions in [Fe/H] for age groups from young to old: 0.11, 0.14, 0.16, 0.17, 0.15, and 0.7. In all but the oldest bin the distributions appear Gaussian. The shape of the red histogram (oldest stars) can be approximated better by a log-normal distribution because of its strong negative skewness. The metallicity peak of this oldest population is around $-0.5$, with a range of values that agree well with the results by \cite{soubiran03} for their thick-disk selection of clump giants. As gathered from Fig.~\ref{fig:r0}, more than 50\% of these old stars were born at $r<5$~kpc. The causes for their outward migration are discussed in Sec.~\ref{sec:liu}.

The bottom row of Fig.~\ref{fig:chem} is the same as the top one, but for the [Fe/H]--[O/Fe] relation. The relationship between [O/Fe] and age, displayed in the rightmost panel, shows the good correlation between the two, assuring that $\alpha$-elements can be a good age-indicator. However, this figure also shows that the correlation is best seen for older ages. For age$\lesssim$6~Gyr, the different distributions overlap significantly (see also Fig.~\ref{fig:sig}, bottom). This should be kept in mind when drawing conclusions from large spectroscopic surveys, where often the [$\alpha$/Fe] ratio is used as a proxy for age.

By correcting for the spectroscopic sampling of stellar subpopulations in the SEGUE survey, \cite{bovy12a} showed recently that the bimodality seen in the [O/Fe]-[Fe/H] distribution for the uncorrected sample disappears. As the middle bottom panel in Fig.~\ref{fig:chem} shows, our results agree in that our simulated, unbiased [O/Fe]-[Fe/H] stellar density distribution is smooth. In Sec.~\ref{sec:sel} we show that biases introduced by the preferential selection of thin- and thick-disk stars, as commonly employed in observations, can result in a gap in the [O/Fe]-[Fe/H] plane.

\subsection{Migration effects on the chemical gradients}
\label{sec:grad}

In Fig.~\ref{fig:grad} we show how the initial chemical gradients are affected by the disk dynamics. We divided the galactic disk into six age-bins and plotted the [Fe/H] and [O/Fe] gradients by lines of different colors, as indicated in the bottom panel. Solid and dotted line-styles represent the initial (chemical) and final (chemodynamical) states, respectively. The bar's CR and OLR (estimated at the end of the simulation) are given by the dotted-red and solid-blue vertical lines, respectively. A strong flattening in [Fe/H] is seen for the older populations, but the younger stars are much less affected. For stars younger than 2~Gyr the final gradient is very similar to the initial one out to $\sim12$~kpc. This is related to the fact that the majority of stars in this age-bin are born near their final radii, as we found out was the case for the ``solar" radius (see dotted black line in Fig.~\ref{fig:r0}, left). Note that the bar's CR acts as a pivot point around which the initial metallicity profiles turns counterclockwise, raising values in the disk and lowering them in the bulge/bar region. This is easy to understand by recalling that the bar CR radius is where the strongest exchange of angular momentum occurs (see Fig.~\ref{fig:xy}, bottom row) and is one example of the importance of considering the effect of the bar in modeling the Galactic disk.

To give a quantitative idea for the change in the metallicity gradient of young stars, we fitted lines to the solid (initial) and dotted (final) black lines in Fig.~\ref{fig:grad}, top panel. We estimated that the original gradient of $\approx-0.061$ or $\approx-0.057$~dex/kpc, depending on the radial range we used for fitting (5-12 or 6-11~kpc, respectively) changes to $\approx-0.059$ or $\approx-0.058$~dex/kpc at the end of the simulation. This small effect on the gradient of the youngest stellar population is reassuring, since this was used as a constraint for our input chemical evolution model.

Incidentally, the CR (which is consistent with what is expected for the MW) is almost exactly at the radius where the input metallicity gradient begins to steepen (Fig.~\ref{fig:grad}, top). While the chemical model we used was not motivated by the effect of the bar and is, in fact, a somewhat extreme case regarding the strong increase in [Fe/H] at $r<5$~kpc, this choice may well be the correct one, considering the stronger chemical enrichment expected inside bars (e.g., \citealt{ellison11}). 

Much weaker flattening is seen for the [O/Fe] profiles, related to the progressively weaker radial variation of older samples. Owing to the reversed gradients compared to [Fe/H], a {\it clockwise} turn is now seen around the CR pivot point. In both panels of Fig.~\ref{fig:grad} the values in the innermost disk region are unaffected. This is because most of these stars are trapped inside the bar, as mentioned above, preventing their escape once the bar is formed.

\begin{figure}
\includegraphics[width=8.cm]{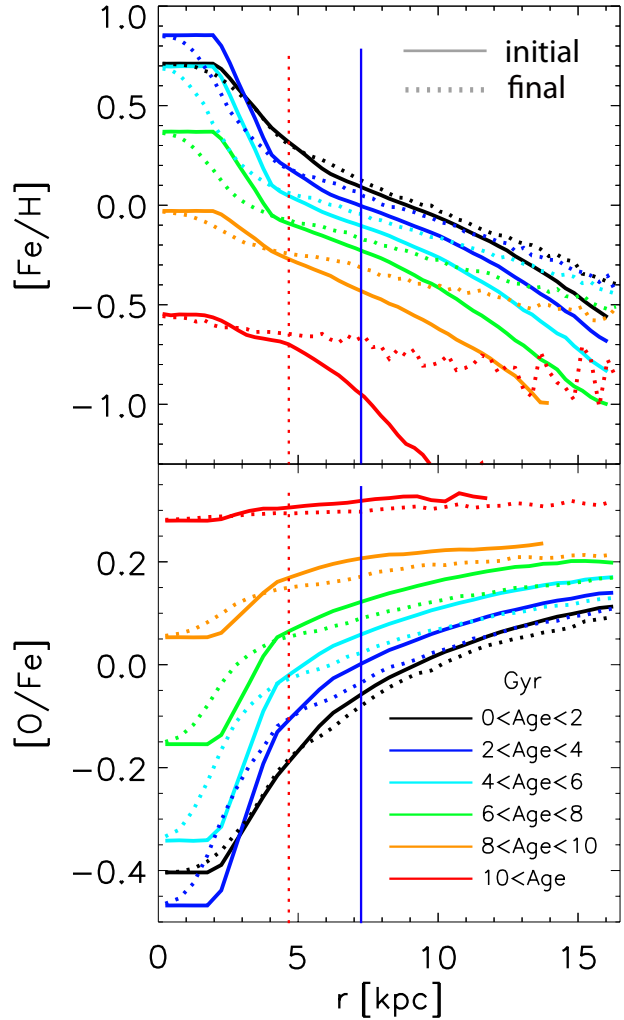}
\caption{
Effect on the initial [Fe/H] (top) and [O/Fe] (bottom) gradients for different stellar age groups. The solid and dotted color curves show the initial and final states, respectively. Note that while strong flattening is observed for the older populations, the metallicity gradient for the youngest stars ($age<2$) is hardly affected at $r\lesssim12$~kpc, thus justifying the use of our chemical model, which uses this as a constraint.
}
\label{fig:grad}      
\end{figure}

\begin{figure}
\includegraphics[width=8.cm]{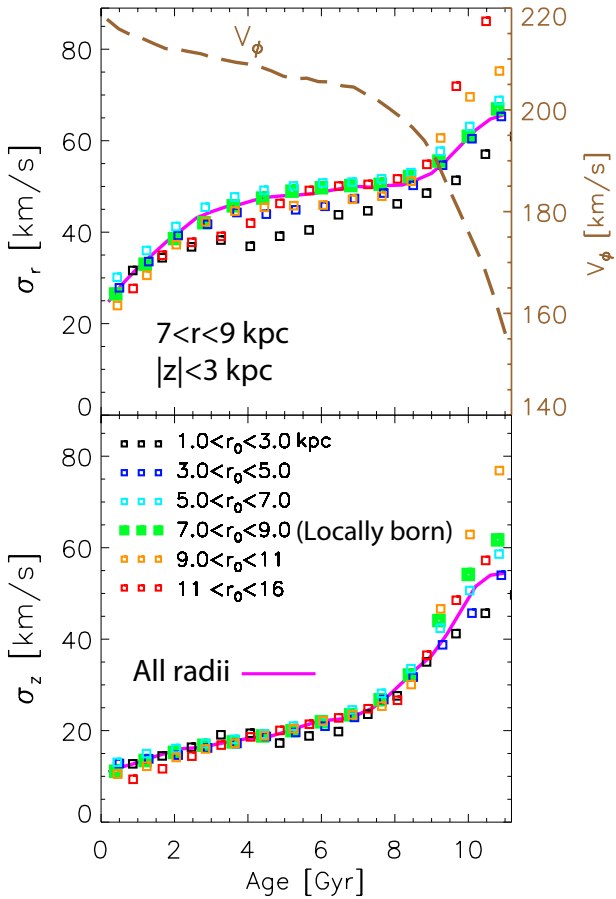}
\caption{
Age-velocity relations resulting from our model in the radial (top) and vertical (bottom) directions for stars in the solar cylinder $7<r<9$~kpc, $|z|<3$~kpc. The pink solid lines show the AVRs estimated from all stars, while color-coded square symbols indicate the contribution from samples born at different distances from the galactic center. The brown line in the top panel shows the stellar rotational velocity, $V_\phi$.
}
\label{fig:sig}      
\end{figure}

\begin{figure}
\includegraphics[width=7cm]{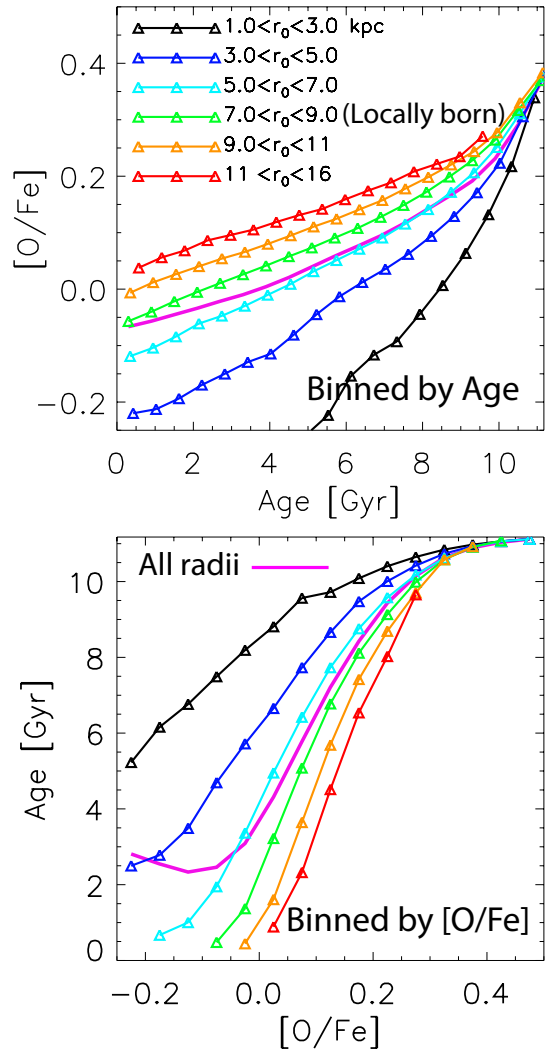}
\caption{
Relation between age and [O/Fe] ratios for our simulated disk. {\bf Top:} Mean [O/Fe] as a function of age for the total sample (pink curve) and for samples born at different galactic radii, as color coded. {\bf Bottom:} Mean age as a function of [O/Fe] ratios.
}
\label{fig:age_ofe}      
\end{figure}

The prediction that the [O/Fe] radial profiles are essentially preserved is an important result and can provide a powerful constraint on different input chemical evolution models. The shapes of the [O/Fe]-profiles for different age bins are a direct consequence of our adopted variation of the infall-law with radius (and hence the SFHs at the different positions). In the near future, these will become possible to measure by combining the good distances and ages expected from the CoRoT mission \citep{baldin06}, with abundance ratios obtained by spectroscopic follow-up surveys (see \citealt{chiappini12,freeman12}). For the young populations, this should be already possible to obtaine from the observations of open clusters, e.g., with the ongoing Gaia-ESO \citep{gilmore12} or APOGEE \citep{majewski10} surveys.

\subsection{Age-Velocity relation}
\label{sec:sig}

An important constraint in MW disk modeling is the observed velocity dispersion in the solar neighborhood. Due to the lack of good age estimates, the shape of the age-velocity relation (AVR) has been a matter of debate. While a power-law of the form $\sigma=t^{0.2-0.5}$ has been proposed (e.g., \citealt{wielen77, dehnen98a, binney00}), some works have predicted an increase for stars younger than 2~Gyr, followed by a saturation up to $\sim$10~Gyr, followed by a strong increase associated with the thick disk, possibly related to strong merger activity (or a single major merger) at high redshift (e.g., \citealt{freeman91,edvardsson93,quillen00}).

In Fig.~\ref{fig:sig} we present the radial (top) and vertical (bottom) velocity dispersions, $\sigma_r$ and $\sigma_z$, as functions of age for stars in the simulated solar cylinder ($7<r<9$~kpc, $|z|<3$~kpc). The solid pink curves plot the total sample, while the color-coded square symbols show stellar samples born in six different radial bins, as indicated in the bottom panel. 

Examining the total population, we note that for ages older than $\sim3$~Gyr, $\sigma_r$ flattens significantly, rises only by $\sim$5~km/s in the time interval 3-8~Gyr, after which the slope increases again, finally reaching $\sim65$~km/s for the oldest stars. A similar behavior is seen for $\sigma_z$, although the knee at 2-3 Gyr is less pronounced. A sharp increase, even more drastic than the one seen for $\sigma_r$, is observed for $\sigma_z$ beyond $\sim8$~Gyr, reaching $\sim55$~km/s for the oldest stars. These trends are consistent with the second scenario mentioned above, favoring a violent origin for the hottest stellar population in the solar neighborhood. 

\subsubsection{Contribution from stars born at different galactic radii}
\label{sec:mig1}

We now compare the contribution to the AVRs from stars born at different galactic radii, as indicated by the color-coded square symbols in the left column of Fig.~\ref{fig:sig}. We first note that the locally born sample (green filled squares) mostly follows the total population (pink curve) and is even hotter for the oldest ages. This is clear evidence that in our simulation migration does not contribute to the heating in the ``solar" vicinity and, in fact, suppresses it for the oldest stars (which comprise the thick disc).

For ages~$<3-4$~Gyr there exists a general trend in both $\sigma_r$ and $\sigma_z$, that stars born at progressively larger radial bins are dynamically cooler -- subsamples born outside/inside our solar bin have slightly lower/higher velocity dispersions. Note that the contribution from hotter outward migrators is suppressed by cooler inward ones, with the result that the total population's velocity dispersions remain the same as the locally born sample. This is consistent with the recent findings by \cite{minchev11b} and \cite{minchev12b}, who demonstrated that migrators are generally not expected to heat the disk because of the approximate conservation of the average radial and vertical actions in the absence of external perturbers. It is remarkable that we do not see any significant contribution here either, although the disk we study is exposed to satellite perturbation until the final time, which probably breaks the action invariance. We note that no massive mergers have occurred in our simulation for the last $\sim8-9$~Gyr, which could explain this behavior. 

The picture changes for ages~$>3-4$~Gyr. A reversal is seen in the general trend, where now samples born at progressively smaller radii are dynamically cooler, not hotter. For example, stars born in the range $1<r_0<3$~kpc (black squares) have between $\sim5$ and $\sim15$ km/s radial velocity dispersion lower than the total (and local) population. In contrast, stars born in the solar vicinity and outward exhibit trends similar to the total population in the range $4\lesssim$~age~$\lesssim9$~Gyr; the velocity dispersions for these stars are much higher for ages~$\gtrsim9$~Gyr. The difference among the velocity dispersion trends for samples born at different radii are more obvious for $\sigma_r$, but the $\sigma_z$ behavior is similar.

\begin{figure*}
\includegraphics[width=18cm]{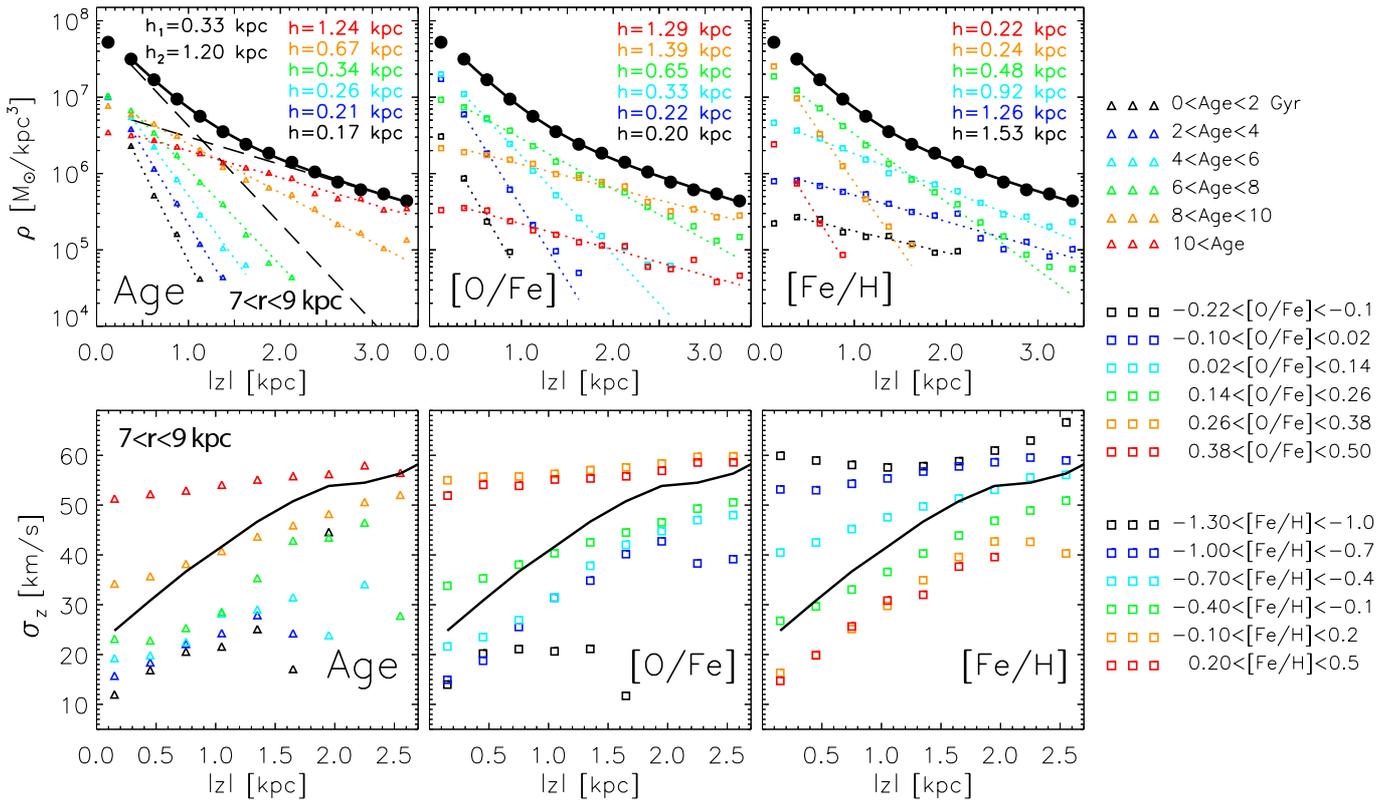}
\caption{
{\bf First row:} Vertical mass density as a function of height (absolute value) above the disk plane, $\rho(|z|)$, for stars in the ``solar" vicinity binned in six different groups by age (left), [O/Fe] (middle), and [Fe/H] (right). The corresponding bin values are shown to the right of the figure. The color-coded values, $h$, indicate single exponential fits shown by the dotted lines. Two exponentials (dashed black lines) are fit to the total density in the left panel with filled circles indicating simulated data and solid line the fit. An abrupt increase in scale-height for the oldest/highest [O/Fe]/most metal-poor two samples can be seen, indicating the end of the merger-dominated early epoch. {\bf Second row:} Same as above, but for the vertical velocity dispersion, $\sigma_z(|z|)$. Very similar increasing trends are seen for young stars, high [O/Fe], and metal-rich bins. As populations age, relations flatten. The black solid line is the total sample.
}
\label{fig:den}      
\end{figure*}

\subsection{Relation between age and [O/Fe] ratios}
\label{sec:ageofe}

We now briefly examine the relation between age and [O/Fe] ratios for stars ending up in our simulated solar vicinity. In the top panel of Fig.~\ref{fig:age_ofe} we plot the mean [O/Fe] as a function of age for six birth-radius bins at the final simulations time. The pink curve shows the total sample, which for ages~$\gtrsim4$~Gyr is seen to follow the subsample originating at $5<r_0<7$~kpc (cyan), rather than the locally born population (green). This agrees well with Fig.~\ref{fig:r0}, left panel, which showed that the $r_0$-distribution peaks at around 6~kpc. For ages~$\lesssim8$~Gyr, approximately linear relations are found for stars born in a given radial bin. However, due to contamination from radial migration, for the total sample the scatter in [O/Fe] increases strongly with decreasing age, in agreement with the bottom-right panel of Fig.~\ref{fig:chem}.

The bottom panel of Fig.~\ref{fig:age_ofe} shows the mean age as a function of [O/Fe] ratios. We see that stars with [O/Fe]$\lesssim-0.1$ have a mixture of ages~$<4$~Gyr. While the total sample (pink curve) follows the stars born at mean radii $r_0\sim6-8$~kpc (green and cyan bins) from high to low [O/Fe] ratios, at [O/Fe]~$\lesssim-0.05$ dex it shows an abrupt  shift toward the $3<r_0<5$~kpc bin. This introduces a strong nonlinearity in the age-[O/Fe] relation for the youngest ages.

In summary, the approximately linear relation between age and [O/Fe] {\it for stars born at the same radius} is destroyed when radial migration is taken into account. When grouped by common birth radii, each sample is a quite good age-indicator in the covered [O/Fe] region, e.g., in the range $-0.1\lesssim$[O/Fe]$\lesssim0.4$ for the locally born stars. However, due to the inevitable radial mixing, the overlap of these subpopulations combined with the offset in the [O/Fe]-range covered contaminates this relation.

\subsection{Vertical structure of the local disc}
\label{sec:vertical}

In this section we study the density and vertical velocity as functions of height above the disk plane in the annulus $7<r<9$~kpc. 

The first row of Fig.~\ref{fig:den} displays the vertical mass density as a function of height (absolute value, $|z|$) above the disk plane. The total density is indicated by the filled black circles in each top panel. Two exponentials are fit to it in the left panel, shown by the dashed black lines, with the solid black line plotting the sum. We obtain very similar values to the expected MW thin and thick disk scale-heights: $h_1=330$~pc and $h_2=1200$~pc, respectively.   

To assess the stellar age and chemical composition that comprise the thin and thick disks, we split the total mass into six different groups by age (triangles, left), [O/Fe] (squares, middle), and [Fe/H] (squares, right). The corresponding bin values are shown to the right of Fig.~\ref{fig:den}. We find that single exponentials can fit each subsample reasonably well by omitting the 200~pc closest to the disk plane. The color-coded values, $h$, indicate the fits, overplotted by the dotted lines. There is a general trend of increase in the scale-heights with increasing age, increasing [O/Fe], and decreasing [Fe/H], running approximately between $\sim0.2$~kpc and $\sim1.3$~kpc (1.53~kpc for the most metal-poor bin). However, we note an abrupt increase in scale-height (by a factor of $\sim2$) for the oldest two and the highest [O/Fe]/most metal-poor three subsamples. 

It is easy to see that this thick-disc-like component is mostly composed of stars with ages~$>8$~Gyr, [O/Fe]~$>0.15$, and [Fe/H]~$<-0.5$. This is the age beyond which the AVRs show a strong increase, which we discussed in Sec.~\ref{sec:sig} (see Fig.~\ref{fig:sig}) and related to the merger-dominated early epoch of our simulation, as well as high stellar birth velocity dispersions. It also became clear in Sec.~\ref{sec:mig1} that the disk does not thicken because of radial migration, but instead migration was found to suppress disk heating. It is clear, therefore, that the thick-disc-like part of the vertical density in our simulated solar vicinity results from merger activity.

In the second row of Fig.~\ref{fig:den} we show the vertical velocity dispersion as a function of height above the disk plane, $\sigma_z(z)$, for stars in the ``solar" vicinity ($7<r<9$~kpc) binned in the same groups as in the first row. Similar increasing trends are seen for young stars, low-[O/Fe] and metal-rich bins. Interestingly, as populations get older, relations flatten around $50-60$~km/s for the oldest/high-[O/Fe]/metal-poor samples. Both these trends and values are similar to the recent observational results by \cite{bovy12c}. We note that in our work subpopulations are in narrow bins of [O/Fe] {\it or} [Fe/H], but not both at the same time, as done by \cite{bovy12c}. Nevertheless, the similarity with their Fig.~2, bottom, is obvious. The solid line in each panel shows the variation of $\sigma_z$ with $|z|$ for the total density. The approximately linear relation flattens at $|z|\gtrsim2$~kpc, where the thick disk starts to dominate the stellar mass.

It is also interesting to consider the variation of [Fe/H] with height above the disk plane, $z$, as well as with mean vertical velocity, $v_z$. In Fig.~\ref{fig:fe_vz} we show these relations for the same bins of age (left) and [O/Fe] (right) used for Fig.~\ref{fig:den}. We considered stars above and below the plane and took the mean absolute values $|z|$ and $|v_z|$. The solid black lines in each panel display the total number of stars in the annulus $7<r<9$~kpc. A negative trend is seen for both $|z|$ and $|v_z|$ as expected from observations (e.g., \citealt{katz11,schlesinger12, rix13}). As in the $|z|$-$\sigma_z$ relation, we find a flattening at $|z|\gtrsim2$~kpc for $|z|$-[Fe/H], as well as in the $|v_z|$-[Fe/H] relation at $|v_z|\gtrsim90$~km/s.

When divided into age or [O/Fe] subsamples, the variation of [Fe/H] is mostly lost for each individual group. This is true for both $|z|$ and $|v_z|$. It is then clear that the downtrend in the total sample comes from the fact that younger (or $\alpha$-young) stars are, on the average, both metal-rich and more numerous at small distances from the plane. With increasing height the density of older (high [O/Fe]-ratios) and thus more metal-poor populations continuously increases, resulting in the overall negative gradient with $|z|$. 

\begin{figure}
\includegraphics[width=8.5cm]{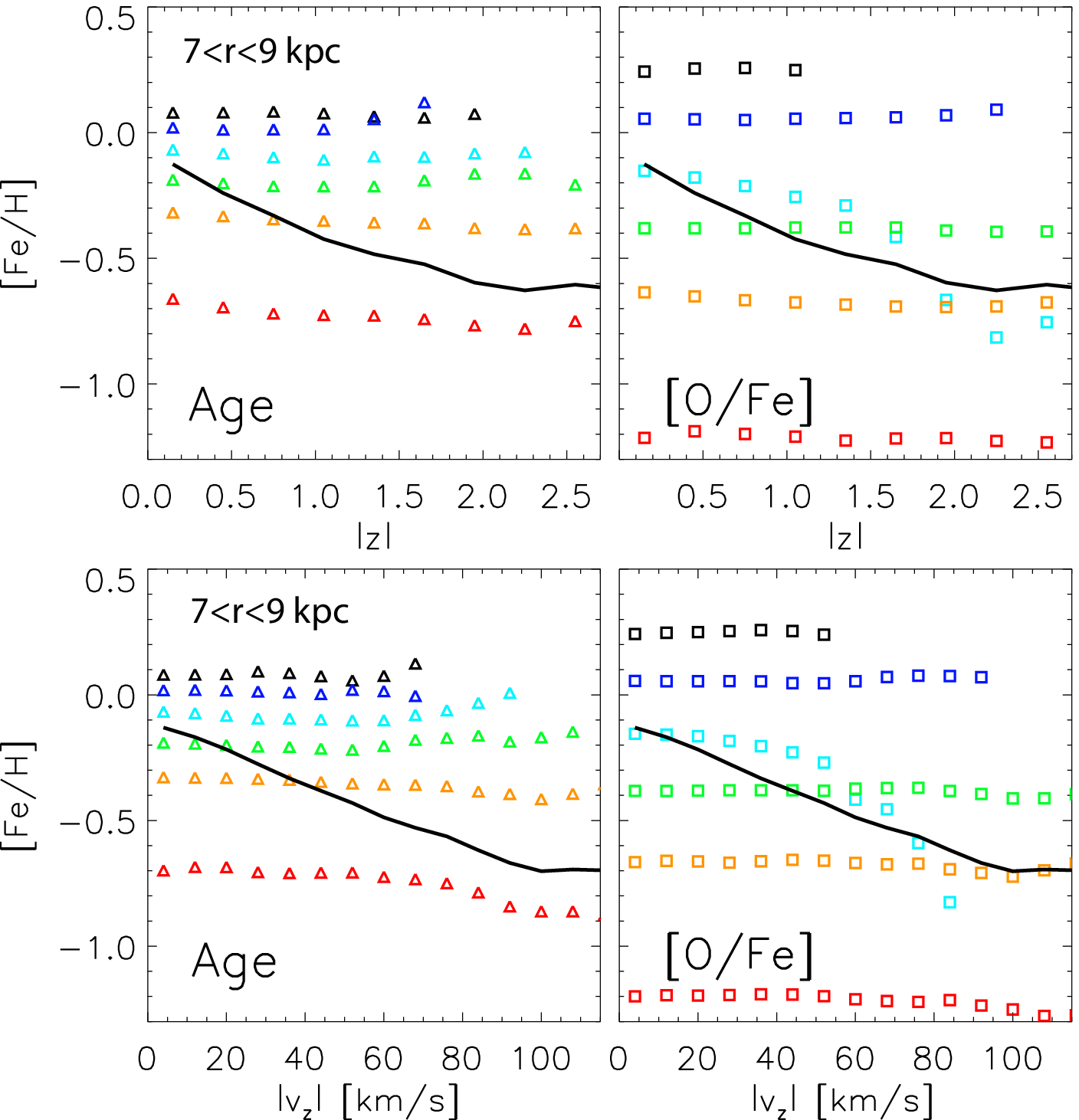}
\caption{
{\bf First row:} [Fe/H] as a function of height above the disk plane (absolute value), $|z|$, for stars in the ``solar" vicinity. Different colors indicate age (left) or [O/Fe] (right) for the same bins as in Fig.~\ref{fig:chem}. {\bf Second row:} Same as above, but the vertical axes show the absolute mean vertical velocity, $|v_z|$. The solid black curve in each panel shows the relation for the total sample.
}
\label{fig:fe_vz}      
\end{figure}

Similarly, younger stars are confined close to the disk and, thus, have relatively low vertical velocities. With increasing age (and thus velocity dispersion) the amplitude of vertical oscillations increases, while the stellar density decreases. This, again, results in the negative trend of [Fe/H] with $|v_z|$. The flattening at $|v_z|\gtrsim90$~km/s (and $|z|\gtrsim2$~kpc) arises because the number of stars with such high velocities (and maximum $|z|$-values achieved) is almost exclusively dominated by the oldest/[O/Fe]-rich stars.

The approximately flat mean $|z|-$[Fe/H] and $|v_z|-$[Fe/H] relations for stars of narrow age- or [O/Fe]-bins is an extension to the similar behavior seen in the $|z|-\sigma_{z,r}$ plots above and provides additional predictions for Galactic surveys. 

\section{Relation to recent observational results}

We have already shown that a number of the basic observed chemistry and kinematics of solar neighborhood stars can be explained by our model. In this section we compare our model predictions directly with recent observational data. To be able to compare our results with different MW surveys we need to spatially constrain our model sample. For example, the GCS and high-resolution stellar samples are confined to within 100~pc from the Sun, while the SEGUE and RAVE samples explore much larger regions, but miss the local stars. 

\begin{figure}
\centerline{\includegraphics[width=8.5cm]{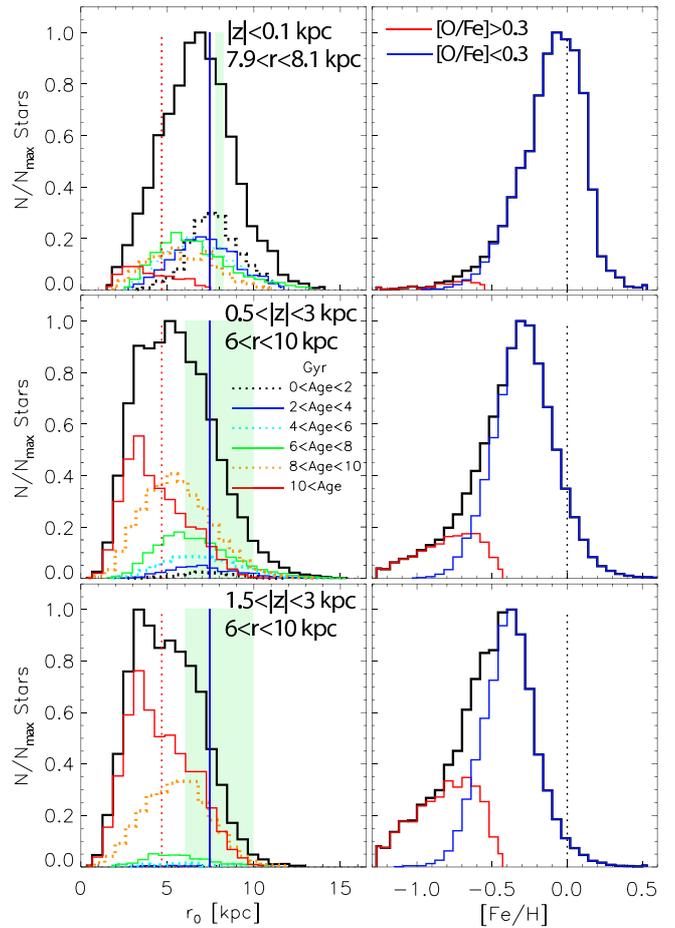}}
\caption{
Changes in the metallicity distribution function (MDF) for samples at different distances from the disk plane. {\bf First column:} Distribution of birth radii for stars ending up in the bin $7<r<9$~kpc (green rectangle) at the final time. The bar CR and OLR (estimated at the final time) are indicated by the dotted red and solid blue vertical lines. From top to bottom, we have applied the selection criteria $|z|<0.5$, $|z|>0.5$, and $|z|>1.5$~kpc. {\bf Second column:} MDF for the same particle subsets as on the left. The black line shows the total population for each selection, while the red and blue colors indicate [O/Fe]$>0.3$ and [O/Fe]$<0.3$, respectively. A shift in the peak from [Fe/H]=0 to negative values is seen with increasing distance from the disk plane.
}
\label{fig:mdf}      
\end{figure}

\subsection{Changes in the MDF for samples at different distances from the disk plane}
\label{sec:segue}

In Fig.~\ref{fig:chem} we presented the metallicity distribution function (MDF) in the solar vicinity as it results from our unbiased model. To see the biases introduced to the MDF when distance cuts are applied, we explore three possibilities below and compare with observations.
 
Fig.~\ref{fig:mdf} shows our predicted MDF at the solar vicinity for different distances from the disk plane. From top to bottom, we applied the selection criteria $7.9<r<8.1$, $|z|<0.1$~kpc (similar to what is seen by GCS); $6<r<10$, $0.5<|z|<3$~kpc (close to SEGUE coverage); and $6<r<10$, $1.5<|z|<3$~kpc. 

The left column of Fig.~\ref{fig:mdf} shows the distribution of birth radii (as the left panel of Fig.~\ref{fig:r0}) for stars ending up in the corresponding radial bin (green-shaded vertical strip) at the final time. The majority of stars are younger than 8~Gyr and only a small fraction comes from inside the bar's CR (red dotted vertical). As the distance above the plane increases, stars with ages~$>8$~Gyr start to dominate the samples, most of them born inside the bar's CR.  

In the right column of Fig.~\ref{fig:mdf} we show the MDF for the same particle subsets as on the left. The black lines show the total population for each selection, while the red and blue colors indicate [O/Fe]~$>0.3$ and [O/Fe]~$<0.3$, respectively. It can be seen that the MDF peak shifts from solar at low heights from the plane (consistent with the GCS sample, e.g., \citealt{casagrande11}), to a value below solar, because we selected only particles with heights above 500~pc (consistent with RAVE -- \citealt{boeche13} and SEGUE -- \citealt{bovy12a, schlesinger12}). Note that even at large distance from the plane, the MDF is dominated mostly by stars with [O/Fe]~$<0.3$, whereas the higher [O/Fe] values are always contributing to its metal-poor tail. This suggests a bias in the distributions obtained by \cite{lee11}, as discussed by \cite{bovy12a}.

\begin{figure*}
\centerline{\includegraphics[width=16.cm]{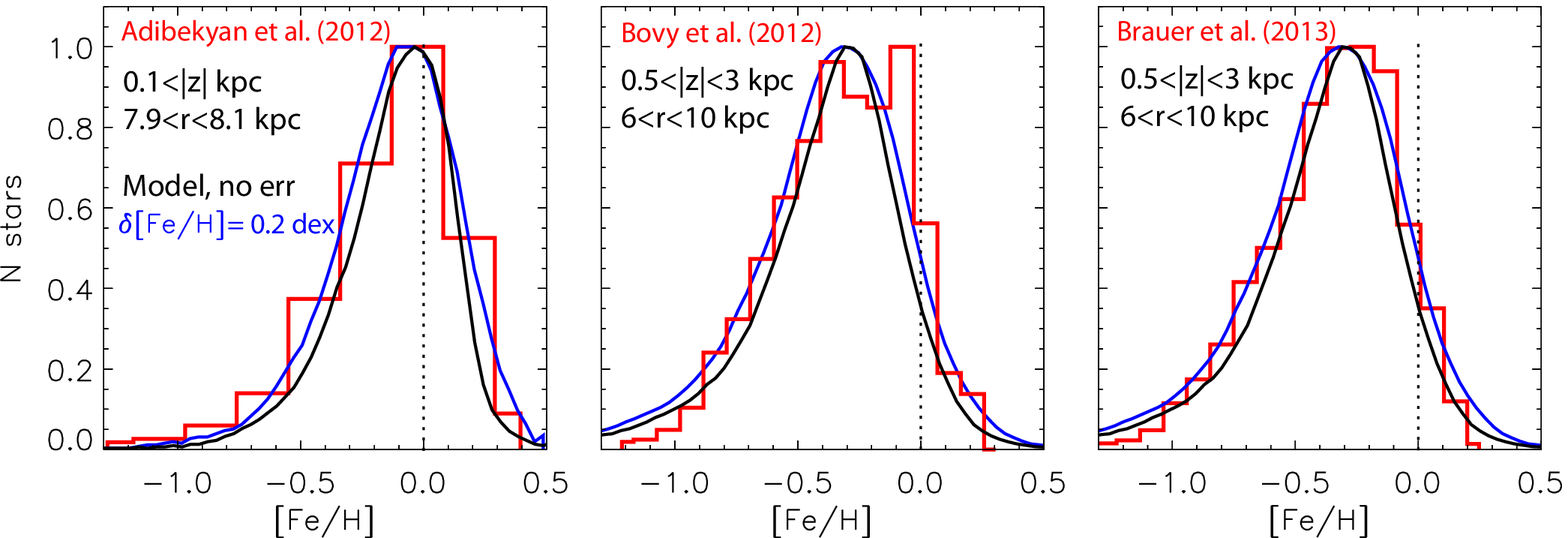}}
\caption{
Comparison of our model metallicity distribution function (MDF) with recent observations. The red histograms shows data from \cite{adibekyan12} (left), \cite{bovy12a} (middle), and Brauer et al. (2013) (right). In each panel the black and blue curves plot the model data with no error in [Fe/H] and convolved error of $\delta$[Fe/H]=0.2~dex, respectively. We have selected a very local sample to compare with the high-resolution nearby sample by \cite{adibekyan12} (as indicated) and a range in $z$ and $r$ to approximately match the SEGUE data studied by the other two works. A shift in the peak from [Fe/H]$\sim$0 to [Fe/H]$\sim-0.3$ dex for both observed and simulated data is seen when considering a higher sample depth. 
}
\label{fig:mdf1}      
\end{figure*}

In Fig.~\ref{fig:mdf1} we finally compare our model MDF with recent observations. We consider the high-resolution sample by \cite{adibekyan12} (histogram extracted from Fig.~16 by \citealt{rix13}), the mass-corrected SEGUE G-dwarf sample by \cite{bovy12a} (MDF extracted from their Fig.~1, right panel). The third data set we considered comes from the latest data release (DR) 9, SEGUE G-dwarf sample, which is also mass-corrected in the same way as the sample of \cite{bovy12a}. The main differences to the data used by \cite{bovy12a} are a different distance estimation (spectroscopic instead of photometric), a selection of G-dwarfs based on DR9, instead of on DR7 photometry, and a signal-to-noise ratio $S/N>20$ instead of $S/N>15$. More details can be found in Brauer et al. (2013, in preparation). The exact same cuts in $r$ and $z$ as in the simulation were applied to the latter sample only, since the other two data sets were not available to us.

The red histogram in each panel of Fig.~\ref{fig:mdf1} shows different observational data as indicated; the additional curves plot the model data with no error in [Fe/H] (black) and convolved errors of $\delta$[Fe/H]=$\pm0.2$~dex (blue), drawn from a uniform distribution. We selected a very local sample to compare our simulated data with the high-resolution nearby data by \cite{adibekyan12} (as in Fig.~\ref{fig:mdf}, top) and a range in $z$ and $r$ to approximately match the SEGUE data studied by the other two works (as in Fig.~\ref{fig:mdf}, middle). A remarkable match is found between our model predictions and the data. The expected broadening in the model distributions is seen, with the inclusion of error in [Fe/H].

\subsection{Is the bimodality in the [Fe/H]-[O/Fe] plane due to selection effects?}
\label{sec:sel}

Starting with the assumption of two distinct entities -- the thin and thick disks --, many surveys have selected stars according to certain kinematic criteria (e.g., \citealt{bensby03, reddy06}). For example, \cite{bensby03} defined a (now widely used) method for preferentially selecting thin- and thick-disk stars with a probability function purely based on kinematics. 

We employed the same technique as \cite{bensby03} and extract a thin- and thick-disk selections from our initially unbiased sample (Fig.~\ref{fig:chem}, bottom-middle panel). To obtain a similar number of thin- and thick-disk stars, as done in most surveys, we randomly down-sampled the thin-disk selection. To match the spacial distribution of high-resolution samples, which are usually confined to small distances from the Sun, we imposed the constraints $7.8<r<8.2$~kpc, $|z|<0.2$~kpc. Our results did not change to any significant degree when this sample depth was decreased or increased by a factor of two.

\begin{figure*}
\centerline{\includegraphics[width=17.5cm]{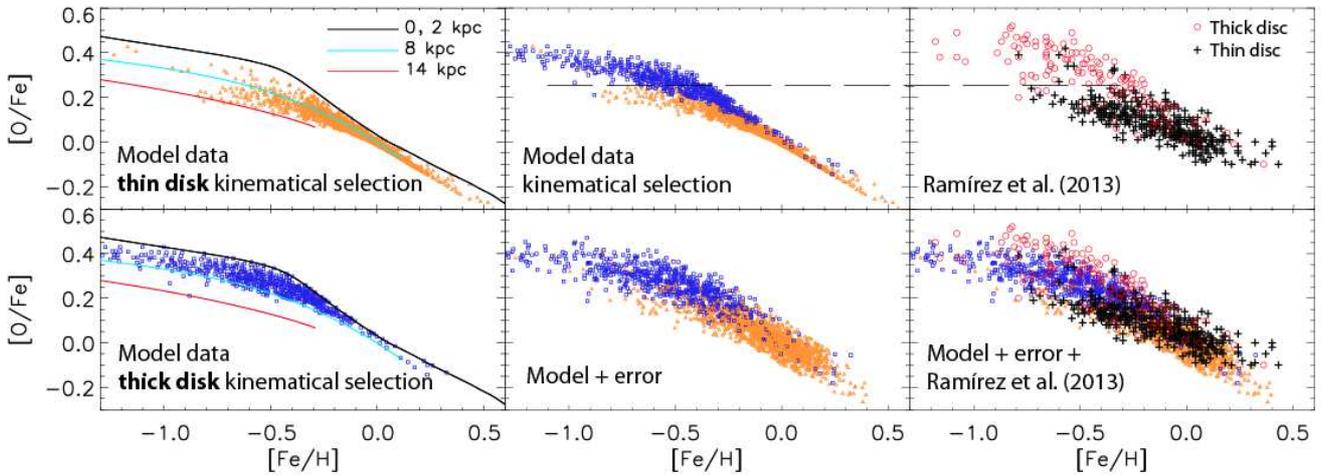}}
\caption{
Selection effects can result in a bimodality in the [Fe/H]-[O/Fe] plane. {\bf Left column:} We have applied the selection criteria used by \cite{bensby03} to extract thin- and thick-disk samples from our model, indicated by orange triangles and blue squares, respectively. {\bf Middle column:} Overlaid thin- and thick-disk model samples, which are shown separately in the left column. The bottom panel shows the same model data but with convolved errors of $\delta$[Fe/H]=$\pm0.1$ and $\delta$[O/Fe]=$\pm0.05$~dex. {\bf Right column:} Top panel plots the high-resolution data from \cite{ramirez13} (their Fig.~11, b). In the bottom panel we overlaid the observed and simulated data samples, which match remarkably well. The black-dashed horizontal line shows the location of the gap, which coincides for the model and data.
}
\label{fig:o_fe1}      
\end{figure*}

The resulting thin- and thick-disk distributions of stars in the [Fe/H]-[O/Fe] plane are presented, respectively, in the top and bottom-left panels of Fig.~\ref{fig:o_fe1}. To assess where stars originate, we overlaid the input chemistry curves for stars born at $r_0\leqslant2$~kpc (black), at $r_0=8$~kpc (or in situ, cyan), and at $r_0=14$~kpc (red). The thick-disk selection appears to originate almost exclusively inside the solar circle (sample is confined between the cyan and black curves). On the other hand, the thin-disk sample follows the local curve for [Fe/H]~$>0.1$ and covers the region between the inner and outer disk (black and red curves) for $-0.4<$~[Fe/H]~$<0.1$. Thin-disk stars with [Fe/H]~$<-0.4$, referred to as the metal-poor thin disc (e.g., \citealt{haywood13}), appear to originate mostly from beyond the solar radius, in agreement with the conclusions by \cite{haywood13}.

The top-middle panel of Fig.~\ref{fig:o_fe1} displays the overlaid thin- and thick-disk selections, which were shown separately in the left column. A relatively clear bimodality is found at [O/Fe]~$\approx0.2$~dex, similarly to what is seen in observational studies. The thick-disk selection follows the upper boundary of the distribution throughout most of the [Fe/H] extent because it originates mostly in the (today's) inner disk. In the bottom-middle panel we convolved errors within our model data of $\delta$[Fe/H]=$\pm0.1$ and $\delta$[O/Fe]=$\pm0.05$ dex, drawing from a uniform distribution.

In the top-right panel of Fig.~\ref{fig:o_fe1} we plot the high-resolution data from \cite{ramirez13} (same as their Fig.~11, b). The gap separating the thin- and thick-disk selections occurs at a very similar location to what is found for the model on the left, as indicated by the black dashed horizontal line. In the bottom-right panel we overlaid the observed and simulated data samples, which match very well. An overall offset within the error ($0.05-0.1$~dex) toward negative [O/Fe]-values is found for the model.

While it is unclear whether the separation between the thin and thick disk in chemistry is real or due to selection effects, the fact that the unbiased volume-completed (although very local) sample of \cite{fuhrmann11} does show a gap argues that it may indeed be the legacy of two discrete stellar populations. We note that after the convolution with errors, the gap between the thin- and thick-disk model selections shown in Fig.~\ref{fig:o_fe1} becomes blurred. This may indicate the need of invoking a discrete thick-disk component, as in the work by \cite{chiappini97}, which will be investigated in a future work of this series.

\subsection{Disk scale-heights of mono-abundance subpopulations}

In Fig.~\ref{fig:den} we showed that although the total vertical mass-density profile in our simulated solar vicinity requires the sum of two exponentials for proper fitting, when grouped in narrow bins by [O/Fe] or in bins by [Fe/H], stellar densities could be fit well by single exponentials. However, using SEGUE G-dwarf data, \cite{bovy12b} showed that the vertical density of stars grouped in narrow bins constrained in both [O/Fe] and [Fe/H] (mono-abundance subpopulations) can also be well approximated by single exponential disk models. 

To see whether we can reproduce these results, in Fig.~\ref{fig:hd} we present the vertical scale-heights of different mono-abundance subpopulations, resulting from our model, as a function of their position in the [Fe/H]-[O/Fe] plane. We convolved errors in our model data of 0.1 and 0.05 dex for [Fe/H] and [O/Fe], respectively. A range in galactic radius of $6<r<10$~kpc, similar to the SEGUE data, was used. Very good fits were obtained by using single exponentials, excluding the 200~pc closest to the disk plane. 

This figure can be directly compared with Fig.~4 of \cite{bovy12b}, bottom-left panel. The general trends found in the SEGUE G-dwarf data are readily reproduced by our model: (i) for a given metallicity bin, scale-heights decrease when moving from low- to high-[O/Fe] populations and (ii) for a given [O/Fe] bin, scale-heights increase with increasing metallicity. Our model predicts scale-heights of up to $\sim1.9$~kpc for the oldest most metal-poor samples; this should be compared with the maximum of $\sim1$~kpc resulting from the mass-corrected SEGUE G-dwarf data of Bovy et al. It is somewhat surprising that for similar local mass-densities and vertical velocity dispersions, the MW disk (as inferred from SEGUE data) appears to be more contracted than our model disk.

\subsection{Mean metallicity as a function of distance from the disc plane}

We showed in the bottom row of Fig.~\ref{fig:fe_vz} that our model recovers the expected negative metallicity gradient as a function of vertical distance from the disk plane, with the prediction that narrow bins in [O/Fe] or age display approximately flat trends. 

We now like compare our model with the recent works by \cite{schlesinger12} and \cite{rix13}, who have used SEGUE G-dwarf data. We extracted the data points from Fig.~15 by \cite{rix13}, where a comparison was made between the vertical metallicity gradient estimated by \cite{schlesinger12} and that implied in the Bovy et al. model \citep{bovy12a,bovy12b}.

Fig.~\ref{fig:fe_z} presents these data along with our model prediction, as indicated by different colors. We find about $0.05-1$~dex higher metallicity than the Bovy et al. model systematically as a function of $|z|$, which results in a better match to the data at $|z|\lesssim1$~kpc. We note that both models (Bovy's and ours) and the data were normalized using the solar abundance values by \cite{asplund05}. It should be kept in mind that while the errors in [Fe/H] are plotted for each data point, distance uncertainties are not shown, but is expected that they should increase with distance from the disk plane. 

A very good agreement between the two models was found for the metallicity variation with $|z|$, including the flattening at $|z|\gtrsim1.5$~kpc. In Sec.~\ref{sec:vertical} this was associated with the predominance of old/high-[O/Fe] stars at large distances from the plane (see Fig.~\ref{fig:fe_vz}).
 
\subsection{Origin of stars with low-metallicity and high-[O/Fe] ratios}
\label{sec:liu}

Analyzing the SEGUE G-dwarf data, \cite{liu12} argued that (internally driven) radial migration cannot be responsible for the origin of the hot metal-poor high-[O/Fe] stars and proposed that they have formed through early-on gas-rich mergers. 

We now search for the origin of the metal-poor high-[O/Fe] stars in our simulation. To approximate the spatial coverage of the SEGUE G-dwarf sample, we considered stars in the annulus $6<r<10$~kpc and distance from the plane $0.5<|z|<3$~kpc. As in \cite{liu12} (their group III), we considered stars with $-1.2<{\rm [Fe/H]}<-0.6$ and $0.25<{\rm [O/Fe]}<0.47$. Finally, we applied a cut in eccentricities at $e=0.2$, thus separating our sample into a cold\footnote{Here we are somewhat stretching the meaning of cold orbits. Usually, for stars at the solar radius, this would be a sample with $\sigma_r\sim5-10$~km/s.} and hot subgroup with corresponding radial velocity dispersions $\sigma_r\sim28$ and 76~km/s. The ratio of cold to hot orbits is $\sim0.33$.

In the top row of Fig.~\ref{fig:liu} we plot stellar density contours in the [Fe/H]-[O/Fe] plane for the hot (left) and cold (right) orbits. As in Fig.~\ref{fig:chem}, we overlay the input chemistry for different initial radii, where the thick cyan curve corresponds to $r=8$~kpc and the rest are separated by two kpc (black for $r=2$~kpc, as in Fig.~\ref{fig:chem}). We can already see that the majority of these stars originates in the inner disk and only a small number comes from the outer disk.

The black histograms in the second row show that the initial radii approximately span the range $1<r_0<10$~kpc for both samples, but the distribution for the cold orbits is much heavier near the solar vicinity. The $r_0$-distribution of these old stars is an example of extremely efficient migration, usually not seen as the effect of internal perturbations. Indeed, as we discussed in Sec.~\ref{sec:sim}, the first couple of Gyr of the disk evolution in our simulation are marked by strong merger activity (see bottom-leftmost panel of Fig.~\ref{fig:xy}). 

\begin{figure}
\centerline{\includegraphics[width=6.5cm]{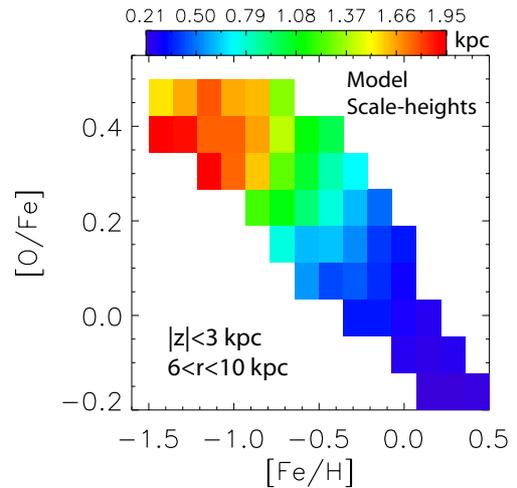}}
\caption{
Model vertical scale-heights for different mono-abundance subpopulations as a function of position in the [Fe/H]-[O/Fe] plane. Very good fits are obtained by using single exponentials, excluding the 200~pc closest to the disk plane. A radial range of $6<r<10$~kpc similar to the SEGUE data is used. This figure can be directly compared with Fig.~4 by \cite{bovy12b}. Our model predicts scale-heights of up to $\sim1.9$~kpc for the oldest, most metal-poor samples; this should be compared to a maximum of $\sim1$~kpc inferred from the Bovy et al. mass-corrected SEGUE G-dwarf data.
}
\label{fig:hd}      
\end{figure}

\begin{figure}
\centerline{\includegraphics[width=6.5cm]{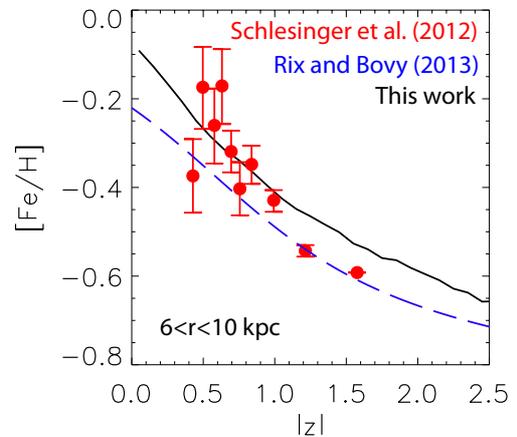}}
\caption{
Mean metallicity as a function of distance from the disk plane, $|z|$, in the range $6<r<10$~kpc. Red filled circles and error bars show SEGUE data from \cite{schlesinger12}, while the blue-dashed line represents the Bovy et al. model. The solid black curve is our model. A very good agreement between the two models in found for the metallicity variation with $|z|$, including the flattening at $|z|\gtrsim1.5$~kpc. In Sec.~\ref{sec:vertical} this was associated with the predominance of old/high-[O/Fe] stars at large distances from the plane (see Fig.~\ref{fig:fe_vz}).
}
\label{fig:fe_z}      
\end{figure}

To see when our hot and cold stellar groups arrive at the solar neighborhood, in the middle row of Fig.~\ref{fig:liu} we plot also the distribution of initial radii obtained 3~Gyr after the beginning of disk formation (red histograms), or about one Gyr after the last massive merger concludes. At this early stage we already see distributions similar to those we find at the final time, e.g., there are three peaks at $\approx3.5$, 5.5, and 7~kpc for both hot samples and the cold selection. Note that not all stars found in the solar vicinity at $t=3$~Gyr would remain there for the next $\sim8$~Gyr due to the continuous migration, although the hot orbits in particular are not expected to migrate very efficiently.  Additionally, part of the cold sample at $t=3$~Gyr would heat up, ending up in the hot one we see today. Support for this expectation is that at the final time the peak at $r_0=7$~kpc for the cold population decreases, while more stars are transferred to the solar vicinity from $r_0=3.5$~kpc as the effect of the bar's CR.

What causes the extreme migration at $t<3$~Gyr? As mentioned in Sec.~\ref{sec:sim}, the last massive merger (1:5 mass ratio) is disrupted by $t\approx2.5$~Gyr, or just before the earlier time considered in the middle row of Fig.~\ref{fig:liu} (red histograms). The strong effect on the changes of stellar angular momenta around this time were presented in Fig.~\ref{fig:xy}, bottom-left panel, leaving no doubt that the early-on merger phase is responsible for the vehement migration we find. 

An additional contribution from the bar in migrating old stars out is expected at $t>3$~Gyr, as inferred from the increase in the $r_0$-distributions at $\sim3.5$~kpc at the final time (bottom row of Fig.~\ref{fig:liu}). It should be kept in mind that migration efficiency decreases with increasing velocity dispersion (e.g., \citealt{comparetta12}), presenting an extreme case in the current kinematic selection. 

\cite{liu12} argued that the old hot metal-poor high-[O/Fe] stars were born hot in gas-rich mergers. Our analysis above suggests a very similar interpretation, where at high redshift stellar samples are both born hot and are additionally heated by mergers. Importantly, this also gives rise to large-scale very efficient migration.

To see when the oldest stars migrate to the solar neighborhood {\it in the lack of early-on massive mergers}, we performed a different model realization for which the dynamics is quiescent and migration is driven mostly by internal perturbers (appendix A, model B). The equivalent of the plots in the second row of Fig.~\ref{fig:liu} for this model B is shown at the bottom row. We find that migration is not as effective at early times (red line), although orbits are less eccentric and the bar is stronger. A sizable fraction of metal-poor high-[O/Fe] stars can still migrate to the solar neighborhood as the effect of the bar. However, this occurs over the entire evolution of the disk and the ratio of cold to hot number of objects is now $\sim0.96$ (to be compared to 0.33 in our standard model), related to the overall low final velocity dispersions. In properly mass-corrected observational data this ratio can be used as a discriminant for the two types of scenarios. A serious deficiency in the merger-free case is that the vertical velocity dispersion of the oldest stars is underestimated by a factor of $\sim2$ (see Fig.~\ref{fig:ap1}, bottom row). 

An interesting result seen in Fig.~\ref{fig:liu}, middle row, is that a significant fraction of stars with thick-disk chemical characteristics are born near the simulated solar vicinity (more than 1/3 in this extended SEGUE-like radial range), i.e., the local thick disc does not come exclusively from the inner disk. This outer disk region is more efficiently heated by mergers than the inner disk, therefore, the hottest disk stars in the solar neighborhood today may have been born in situ and in the outer disc\footnote{Some vertical contraction is expected as the thin disk builds up from gas accretion.} (see also Fig.~\ref{fig:sig}).

\begin{figure}
\centerline{\includegraphics[width=8.5cm]{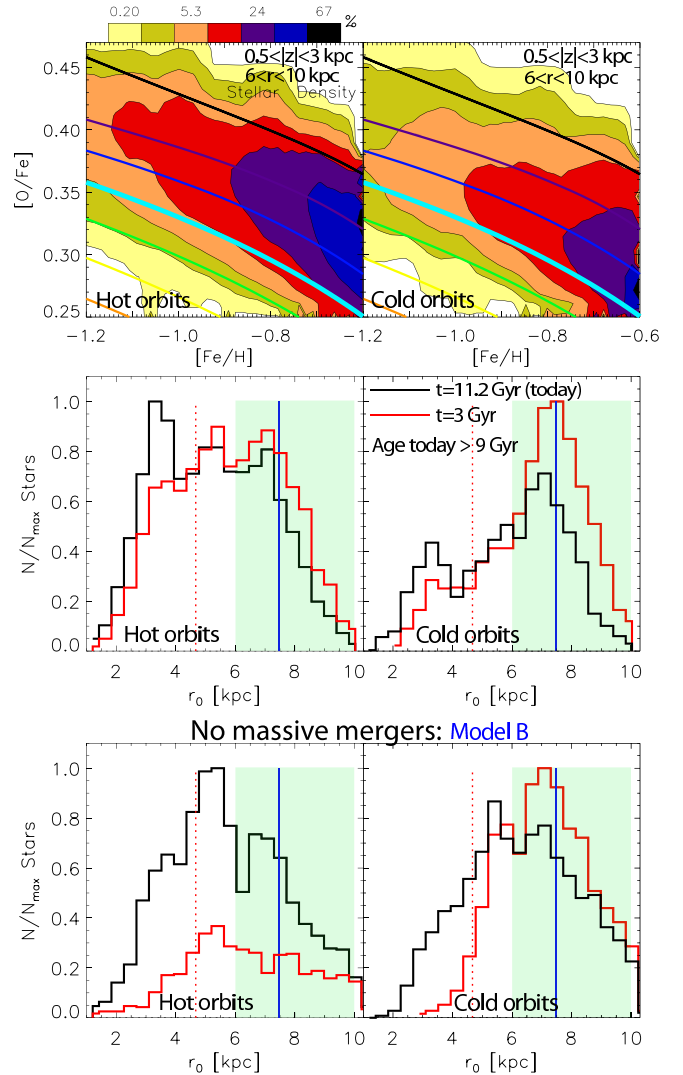}}
\caption{
{\bf Top row:} [Fe/H]-[O/Fe] stellar density distributions for metal-poor high-[O/Fe] stars ending up in the cylinder $6<r<10$~kpc, $0.5<|z|<3$~kpc (similar to SEGUE coverage). Stars are separated into a cold and a hot sample (by a cut in eccentricity at 0.2) with corresponding radial velocity dispersions $\sigma_r\approx28$ and $\sigma_r\approx76$~km/s. Overlaid curves show the input chemistry for different radii as in Fig.~\ref{fig:chem}. {\bf Middle row:} Birth-radius distributions at the final time (black) and at 3~Gyr after the disk starts forming (red) for the selections in the first row. The green strip indicates the radial range at the final time, the dotted-red and solid-blue vertical lines show the bar's CR and OLR positions at the final time. Most stars are in place at the end of the early-on merger phase, with a large fraction born near and outside the solar radius. {\bf Bottom row:} Same as the middle row, but for a model avoiding mergers at high redshift (appendix A). Migration is not as effective at early times (red line), although orbits are less eccentric and the bar is stronger. The fraction of cold to hot orbits is 0.96 instead of 0.33.
}
\label{fig:liu}      
\end{figure}

\section{Unifying model for the Milky Way thick disc}
\label{sec:thick}

We now focus our attention on the properties of the oldest stars (ages~$\gtrsim10$~Gyr) in our model. This subsample is marked by

\begin{itemize}
    \item a metallicity distribution that peaks at [Fe/H]$\sim-0.5$ and has a metal-poor tail down to [Fe/H]$\sim-1.3$ (Fig.~\ref{fig:chem}, upper right panel).
    \item ${\rm [O/Fe]}$-values spanning the range $0.2-0.4$, with a peak around $0.3$ (Fig~\ref{fig:chem}, lower right panel).
    \item a lag in the rotational velocity by $\sim$50~km/s compared to the young stars (Fig.~\ref{fig:sig}, top row).
    \item high velocity dispersions (Fig.~\ref{fig:sig}).
    \item a large scale-height (Fig.~\ref{fig:den}, bottom row).
\end{itemize}

All these properties are strikingly reminiscent of what is referred to as the thick disc in the solar neighborhood, although we used a pure thin-disk chemistry. Within the framework of the model we present here, the MW thick disk has emerged from (i) stars born hot and heated by mergers at early times and (ii) radial migration (from mergers at early times and bar/spirals later on), transporting these old stars from the inner disk to the solar vicinity. A significant fraction of old hot stars today could have been born near to and beyond the solar radius (Fig.~\ref{fig:liu}, middle row).

This conclusion agrees with most (seemingly contradicting) models of thick-disk formation, which expect contribution from only/mostly one of the following: (i) mergers, (ii) early formation in gas-rich turbulent clumpy disks or gas rich mergers, and (iii) radial migration driven by internal instabilities. A combination of these mechanisms working together is required, where strong heating and migration occurs early on from external perturbations (our case) and/or turbulent gas clumps, followed by radial migration taking over the disk dynamics at later times. Yes, mergers are important, but radial migration is also necessary (and unavoidable if a bar, spiral structure and/or mergers are present) to transport out old hot stars with thick-disk chemical characteristics. We note that the thick disk is mostly in place at the time of the last massive merger (Fig.~\ref{fig:liu}, middle row) due to merger-associated larger-scale migration. Yes, migration is important, but the old stars need to be born hot and/or be heated by mergers at high redshift (also unavoidable from our current understanding of cosmology).

The high stellar birth velocity dispersions at high redshift we find in our simulation ($\sim$50~km/s) is consistent with recent works \citep{forbes12, brook12, bird13}. An important dynamical consequence of this is that the disk becomes less susceptible to satellite perturbations (common at high redshift), making it easier for it to survive until today. 

Radial gradients for different distances from the disk plane, as well as the scale-length of this thick disc population will be discussed in detail in paper II.

\section{Comparison to the Sch{\"o}nrich \& Binney (2009a,b) model}

SB09a,b have constructed a model for chemical evolution of the MW that was the first to include radial mixing. The authors have assumed a certain migration efficiency and chemical enrichment scheme in order to fit the current ISM gradient in the MW, the MDF and stellar velocity dispersions. 

Before we compare to our results, we outline two important simplifications in the disk dynamics considered in the SB09a,b model: (i) the presence of a central bar (as found in the MW) was not considered and (ii) the vertical energy (or equivalently, the vertical velocity dispersion) of migrating samples was assumed to be conserved. 

(i) Although SB09a,b have stated that they did not include the effect of the MW bar, their parametrized migration efficiency extends well inside the inner disk, thus requiring the presence of spiral structure in that region. However, it is difficult for spiral instabilities to be supported inside the bulge region due to the high velocity dispersions there. For example, Fig.~15 by \cite{roskar12} contrasts the Fourier amplitudes for barred and unbarred disks: much stronger perturbations are found in the inner disk (r=2 and 4 kpc, black and blue curves, respectively) for their barred simulation, S4, compared with the simulations without bars. The lack of strong effects on the changes of angular momentum at 1-1.5 scale-lengths (near the bar's CR) for non-barred numerical disks is also apparent in Figs. 5 and 15 by \cite{solway12}, and Figures 4, 11, and 12 by \cite{sellwood02}, for example. Despite the lack of explicit inclusion of a bar in the SB09a,b model, the migration efficiency these authors have considered extends to $r<2$~kpc (see Fig.~3 by SB09a and Fig.~1 by SB09b), thus bringing out stars to the outer disk, just as we find in our barred simulation (see Fig.~\ref{fig:r0}). Therefore, the model by SB09a,b effectively emulates the effect of a bar.

(ii) By assuming constancy of the vertical velocity dispersion of migrating populations, SB09a,b were able to successfully reproduce observed chemical trends in the vertical direction (e.g., Fig.~15 by SB09a and Fig.~7 by SB09b). However, as discussed in Sec.~\ref{sec:mig}, in the absence of massive mergers disk thickening due to migration is insignificant: owing to the conservation of vertical action, only extreme migrators contribute by contracting the inner disk and thickening the outskirts. This deficiency in the SB09a,b model affects their main conclusion that a thick disk can be formed in a merger-free MW disk evolution.

We now outline the similarities between our results and those found by SB09a,b:

(1) Migration creates locally a metal-rich tail to the MDF. Note that the metal-rich tail in SB09a,b results from their strong gradient and strong migration efficiency in the inner disk. While the latter is not expected for a disk lacking a bar, as discussed in (i) above, its inclusion into the SB09a,b analytical model works as the effect of the bar. 

(2) Migration introduces a ridge in the [Fe/H]-[O-Fe] plane that becomes a bimodality when combined with the\cite{bensby03} kinematic selection. We note that in Fig.~8 by SB09b, where their model is compared to observations, it is seen that the model's highest [O/Fe] values are higher by about 3~dex than the data shown in the bottom panel. In contrast, our Fig.~\ref{fig:o_fe1} shows a much closer match to the observations. 

The main differences between our model and that by SB09a,b are as follows:

(1) SB09a,b assumed a disk scale-length of 2.5~kpc, which remains constant in time as the disk grows. As a consequence, the oldest disk, which should be consistent with the thick disk, has a final scale-length of 3.1~kpc (their Fig.~7). This is in contrast to our model, where the disk grows inside-out and older disks have increasingly shorter scale-lengths (paper II), consistent with the Segue G-dwarf data analysis by \cite{bovy12a}.

(2) We neglected radial gas flows and the SN-driven wind, possibly resulting in flatter abundance gradients than we currently find. We note that there was no clear justification for the parametrization of these processes by SB09a,b.

(3) Whereas SB09a,b adopted the level of heating required to keep the scale-height of the current thin disk radius-independent and locally corrected, and fitted the parameters that controlled gas flow and the intensity of migration to the GCS MDF, in this work the heating and migration efficiency were drawn from a state-of-the-art cosmological re-simulation, where an important contribution is made by massive mergers at the early stages of disk formation.

(4) The response of stellar vertical motions to migration are more accurately given in the present work, as discussed in (ii) above. 

\section{Conclusions}
\label{sec:concl}

In this work we have presented a new approach for studying the chemodynamical evolution of galactic disks, with special emphasis on the Milky Way (MW). Unlike other similar studies, where either the dynamics was too simplistic or the star formation history (SFH) and chemical enrichment was unconstrained, our chemodynamical model is a fusion between a pure chemical evolution model and a high-resolution simulation in the cosmological context. As we argued in Sec.~\ref{sec:tech}, this new approach allows us to bypass most known problems encountered in fully self-consistent simulations, where chemical enrichment still proves to be a challenge (see Sec.~\ref{sec:cos}). Moreover, this is the first time that a chemodynamical model has the extra constraint of defining a realistic solar vicinity also in terms of dynamics (see Sec.~\ref{sec:sim}).
 
The main results of our chemodynamical model can be summarized as follows:

$\bullet$ 
The distribution of birth radii, $r_0$, of stars ending up in a properly defined solar neighborhood-like location after 11.2~Gyr of evolution peaks close to $r_0=6$~kpc due to radial migration (left panel of Fig.~\ref{fig:r0}). The strongest changes in stellar guiding radii were found for the oldest stars, related to the strong merger activity at high redshift in our simulation and the effect of the bar at later times. Locally born stars of all ages can be found in the solar neighborhood. While a wide range of birth radii is seen for different age groups, the majority of the youngest stars are born at, or close to, the solar neighborhood bin. 

$\bullet$ While the low-end in our simulated solar neighborhood metallicity distribution function (MDF) is composed of stars with a wide range of birth radii, the tail at higher metallicities ($0.25<$[Fe/H]$<0.6$) results almost exclusively from stars with $3<r_0<5$~kpc (see Fig.~\ref{fig:r0}, middle panel). This is the region just inside the bar's CR, which is where the strongest outward radial migration occurs, as we discussed in Sec.~\ref{sec:sim} and showed in the bottom row of Fig.~\ref{fig:xy} (see also Fig.~\ref{fig:ap1} and associated discussion). The fraction of stars in this tail can, therefore, be related to the bar's dynamical properties, such as its strength, pattern speed, and time evolution/formation.
 
$\bullet$
Constraining our simulated sample spatially, we provided excellent matches to the metallicity distribution of stars confined close to the Sun (data from \citealt{adibekyan12}) and the SEGUE G-dwarfs (\citealt{bovy12b} and Brauer et al. 2013), which covers approximately $0.5<|z|<3$~kpc, predicting the expected shift in the distribution peak from solar (nearby sample) to [Fe/H]~$\sim-0.3$ for the SEGUE coverage.
 
$\bullet$  
Our results suggest that the most likely birth location for the Sun is in the range $4.4<r_0<7.7$~kpc, with the highest probability $\sim$5.6~kpc, followed by $\sim$7~kpc (Fig.~\ref{fig:r0}, right). This estimate comes from both dynamical and chemical constraints and is dependent on the migration efficiency in our simulation and the adopted chemical evolution model.

$\bullet$
Examining the effect on the age-metallicity relation (AMR), we found that some flattening is observed, mostly for ages $\gtrsim9$~Gyr (Fig.~\ref{fig:chem}). Although significant radial mixing is present, the slope in the AMR is only weakly affected. This is related to the intermediate distance of the Sun from the Galactic center, where the effect of stars arriving from the inner and outer disks is well balanced, while the migration efficiency is exposed by the scatter around the mean. Therefore, large scatter does not necessarily imply flattening.

$\bullet$ 
We found no bimodality in the [Fe/H]-[O/Fe] stellar density distribution. However, when selecting particles according to kinematic criteria used in high-resolution samples to define thin and thick disks, we recover the observed discontinuity (Fig.~\ref{fig:o_fe1}). This agrees with the recent observational results by \cite{bovy12a}, where a smooth [Fe/H]-[O/Fe] distribution was obtained after correcting for the spectroscopic sampling of stellar subpopulations in the SEGUE survey. A very good match was achieved between our model and the recent high-resolution data by \cite{ramirez13}.

$\bullet$ 
The vertical mass density in our simulated solar vicinity is well fit by the sum of two exponentials giving thin- and thick-disk scale-heights $h_1=330$~pc and $h_2=1200$~pc, respectively (Fig.~\ref{fig:den}, top). By separating this sample into narrow bins of age, [O/Fe], and [Fe/H], we found that the vertical scale-height of each component can be fitted well by a single exponential, with values growing with increasing age, increasing [O/Fe], and decreasing metallicity. An abrupt increase in scale-height was found for samples with age~$\gtrsim8$~Gyr, [O/Fe]~$\gtrsim0.15$, and [Fe/H]~$\lesssim0.5$, related to stars born with high velocity dispersions and strong merger activity at high redshift. The vertical velocity dispersions for each of these subpopulation was found to exhibit smooth variations with height above the disk plane, strongly flattening for old high-[O/Fe] metal-poor stars (Fig.~\ref{fig:den}, bottom). 

$\bullet$ 
A good match was achieved between the mean metallicity variation with height above the disk plane resulting from our model and the data by \cite{schlesinger12} (Fig.~\ref{fig:fe_z}). A good agreement between our model prediction and the model by \cite{bovy12a,bovy12b} was found for the metallicity variation with $|z|$, including a flattening at $|z|\gtrsim1.5$~kpc. This was associated with the predominance of old/high-[O/Fe] stars at large distances from the plane (see Fig.~\ref{fig:fe_vz}).
The metallicity gradients in the $|z|-$[Fe/H] and $|v_z|-$[Fe/H] relations mostly disappear for stars grouped in narrow age- or [O/Fe]-bins (Fig.~\ref{fig:fe_vz}). This provides additional predictions for Galactic surveys. 

$\bullet$ 
For stars in the range $3<$~age~$<8$~Gyr, our model predicts a mostly flat age-velocity relation (AVR) for the radial component, $\sigma_r$, and a weak increase for the vertical component, $\sigma_z$ (Fig.~\ref{fig:sig}). For ages older than 8~Gyr a continuous strong increase is observed for both components, which is related to the massive mergers and stars born with high velocity dispersions at this early epoch. These hot stars can be associated with the MW thick disk. The vertical velocity dispersion of the oldest stars in the absence of mergers (and/or stars born with high velocity dispersions) is underestimated by a factor of $\sim2$, suggesting a violent origin for the MW thick disk (see appendix A, Fig.~\ref{fig:ap1} bottom row).

$\bullet$ We found a strong flattening in the [Fe/H] radial profiles for the older populations with the effect diminishing strongly for the younger ones (see Fig.~\ref{fig:grad}). Stars younger than 2~Gyr have a final gradient very similar to the initial one out to $\sim12$~kpc, justifying its use as a constraint for our chemical model.

$\bullet$ 
We predict that the [O/Fe] radial profiles are essentially preserved for the chemical model we used. The [O/Fe] profiles for different age groups result straightforwardly from the adopted variation of the infall-law with radius (and hence the SFHs at different positions) and, thus, provide a way to constrain different chemical evolution models. In the near future, these will be able to be measured by combining the good distances and ages expected from the CoRoT mission \citep{baldin06}, with abundance ratios obtained by spectroscopic follow-up surveys. For the young populations, this should be already possible to obtain from the observations of open clusters, e.g., with the ongoing Gaia-ESO or APOGEE surveys.

$\bullet$ Finally, probably one of the most important outcomes of our chemodynamical model is that although we used only a thin-disk chemical evolution model, the oldest stars that are now in the solar vicinity show several of the properties usually attributed to the Galactic thick disk. According to the results of the present work, the MW thick disc has emerged naturally from stars migrating from the inner disk very early on due to strong merger activity, followed by additional radial migration driven by the bar and spirals at later times (see Sec.~\ref{sec:liu}). Importantly, a significant fraction of old stars with thick-disk characteristics could have been born near to and beyond the solar radius (Fig.~\ref{fig:liu}, middle row).

We showed that hot old metal-poor high-[$\alpha$/Fe] stars can be delivered to the solar neighborhood at high redshift as the effect of massive perturbers, such as the one with 1:5 mass ratio described in this work (see Fig.~\ref{fig:liu}). For a MW-size galaxy, the occurrence of mergers of this size around $z\sim1$ is consistent with the numerical results by a number of groups (e.g., \citealt{benson04, stoehr06, kazantzidis08, delucia08, stewart08, villalobos08, purcell09}).

Alternatively, a bar can also bring some fraction of these stars to the solar neighborhood, but over a long period of time, as we showed in Fig.~\ref{fig:liu}, bottom row. However, a difference in the metallicity and velocity distributions of these old stars would result. Most notably, avoiding a merger not only avoids the extreme migration at high redshift, but also does not result in high enough velocity dispersions today. For example, we found AVRs with maxima at $\sigma_r\approx50$~km/s and $\sigma_z\approx28$~km/s in the lack of strong mergers, to be compared to $\sigma_r\approx65$ and $\sigma_z\approx55$~km/s for our standard model. Additionally, the ratio of cold to hot orbits (samples separated by an eccentricity cut at 0.2) changes from $\sim$0.33 to $\sim0.96$ when the merger is avoided. This large difference can be used as a constraint in properly mass-corrected observational data to assess the possibility of an early-on merger in the MW.

It is therefore tempting to conclude that the highest velocity dispersion stars observed in the solar neighborhood are a clear indication of an early-on merger-dominated epoch, or equivalently, scenarios involving stars born with high velocity dispersions. Our results suggest that the MW thick disk cannot be explained by a merger-free disk evolution as proposed previously. This is related to the inability of internally driven radial migration to thicken galactic disks \citep{minchev12b}. However, an observational test involving both chemical and kinematic information must be devised to ascertain the effect of early mergers.

While here we concentrated on the solar vicinity, results for the entire disk will be presented in paper II of this series. Chemodynamical predictions for different disk radii may help further constrain thick-disk formation scenarios and the MW assembly in general. 

The results of this work present an improvement over previous models as we used a state-of-the-art simulation of a disk formation to extract self-consistent dynamics and fuse them with a chemical model tailored for the MW. Future work should consider implementing gas flows, Galactic winds, and Galactic fountains in the model to asses the importance of these processes. A suite of different chemical evolution models assigned to different disk dynamics (from different simulations) should be investigated to make progress in the field of Galactic Archeology. Another next step is providing model predictions for the MW disk properties at different Galactocentric regions and for a large number of chemical elements (especially in view of the ongoing high-resolution surveys, such as HERMES and APOGEE).

The method presented here is not only applicable to the MW, but is also potentially very useful for extragalactic surveys. The same approach can be used for any other galaxy for which both kinematic and chemical information is available. A large such sample is currently becoming available from the ongoing CALIFA survey \citep{sanchez12}. Given the SFH and morphology of each galaxy, a chemodynamical evolution model tailored to each individual object can be developed, thus providing a state-of-the-art database for exploring the formation and evolution of galactic disks.

\acknowledgements
We thank the referee for invaluable suggestion that have greatly improved the manuscript. We are grateful to D. Pfenniger and G. Cescutti for useful comments. We thank D.~Brauer, I.~Ramirez,  and  J.~Bovy for providing the data used in Figures~\ref{fig:mdf1}, \ref{fig:o_fe1}, and \ref{fig:fe_z}, respectively.

\begin{appendix}

\section{Effect of different model realizations}
\label{sec:ap1}

\begin{figure*}
\centerline{\includegraphics[width=14cm]{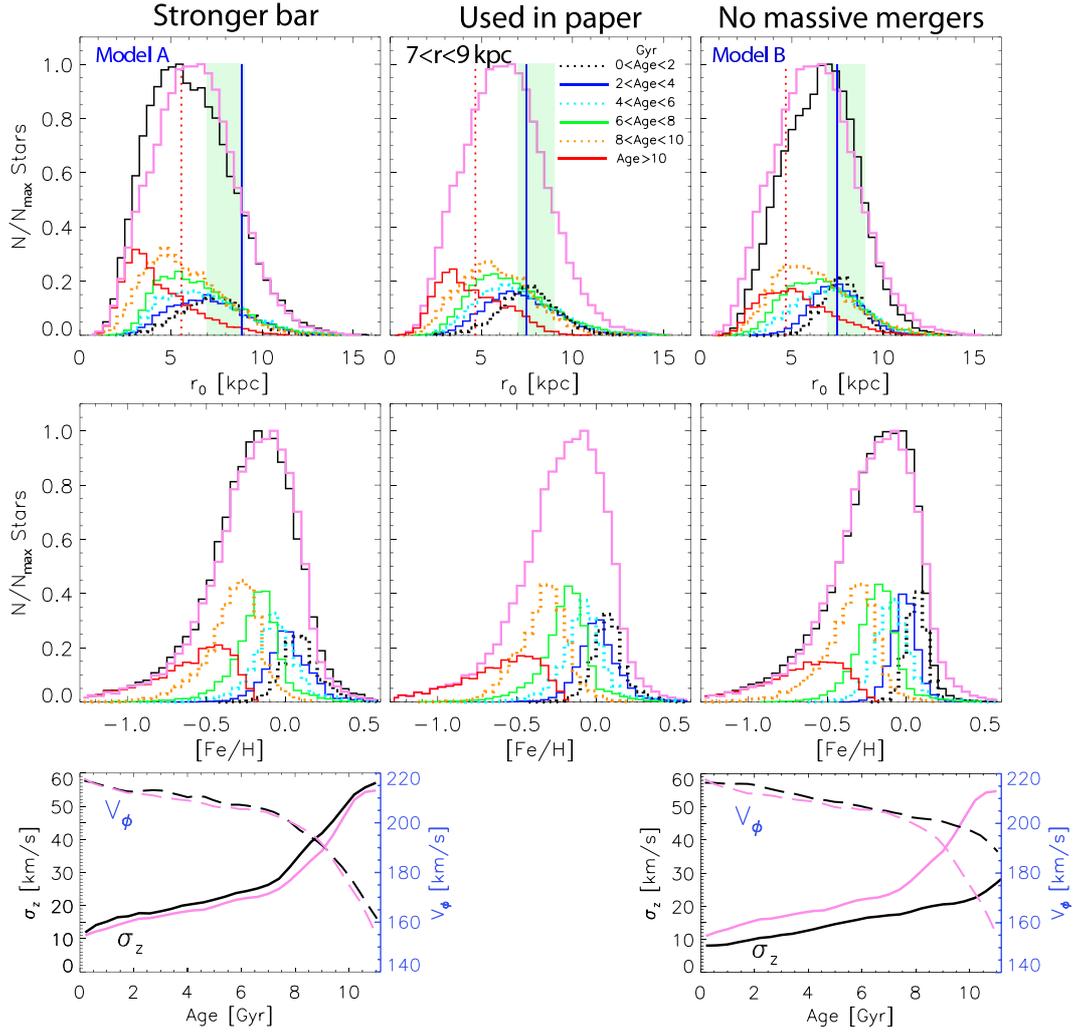}}
\caption{
Comparison between different model realizations. See text.
}
\label{fig:ap1}      
\end{figure*}

As described in Sec.~\ref{sec:sim}, we downscaled our simulated disk by a factor of f=1.67, to place the bar's OLR just inside the solar circle, at $\sim7.5$~kpc (assuming the Sun is at 8~kpc), in agreement with its expected effect on the local disk dynamics (e.g., \citealt{dehnen00, minchev10}). 

We now would like to see how our results change when we rescale the disk by a different factor. We used f=1.4, which places the solar radius $\sim1.25$~kpc inward of the standard location used in the paper. This is equivalent to having a stronger (longer) bar. We refer to this realization as model A.

As a second test, we started the implementation with chemistry 2.7~Gyr later, thus avoiding the early-on massive mergers. We integrated the simulation for additional 2.7~Gyr so that we, again, have 11.2~Gyr of evolution. To keep the correct location with respect to the bar's resonances, we downscaled the disk radius by f=2.1 to account for the bar's slowing down. This realization is referred to as model B.

Note that in the two cases above we only changed the dynamics but not the input chemistry, i.e., we resampled the SFH according to our chemical model, with the 8~kpc chemistry assigned to the rescaled 8~kpc radius in the simulation, etc.

The results are shown in Fig.~\ref{fig:ap1}, where in the first two rows we compare the birth-radius and metallicity distributions resulting from our new realizations model A (left) and model B (right), with our standard model (middle). As expected, when the solar radius is shifted closer to the bar (model A), for all age-bins larger fractions of stars arrive from inside the bar's CR (dotted-red vertical line in the top row); to see this more easily, we overlaid the original total histogram (in pink) in the left and right panels. 

Inspecting the metallicity distributions in the second row of Fig.~\ref{fig:ap1}, we see a shift in the peak of less than $0.1$~dex to negative values when the Sun is closer to the bar. However, the changes are not drastic and the rest of the results in the paper are not affected much by this change. Whether this is true for the entire disk will be investigated in paper II. 

As became evident in the discussion of the age-metallicity relation (Sec.~\ref{sec:amr}), in our simulation we have a deficiency of stars with ages~$>\sim9.5$~Gyr at $r>10$~kpc for our standard model. However, this is a very small fraction of particles, as can be inferred from the bottom-right panel of Fig.~\ref{fig:sfh_sn} (the SFRs at these times and radii attain a maximum at $\sim2$~M$\odot$~pc$^{-2}$~Gyr$^{-1}$ for $r=10$~kpc, sharply decreasing for older stars and larger radii). By shifting the solar radius inward, this artifact is now lost. That this is indeed a very small fraction can be seen by comparing the right tails of the oldest $r_0$-distributions (red histograms) in the first row, left and middle panels. 

The right column of Fig.~\ref{fig:ap1} shows model B, where we avoided the early-on massive mergers phase, but kept the bar resonances at the same distances from the Sun. A smaller number of stars arrives from inside the bar's CR in this case, than in our standard model (see overlaid pink line). This results mostly from the lack of the strong peak in the $r_0$-distribution of the oldest stars (red line). We showed in Sec.~\ref{sec:liu} that the enhancement in the peak at $r_0\sim3$~kpc is due to the massive mergers, which we avoided here.

In the bottom-left and right panels of Fig.~\ref{fig:ap1} we compare the vertical velocity dispersion, $\sigma_z(r)$ (black solid), and the mean rotational velocity, $V_\phi(r)$ (blue dashed), radial profiles resulting from the new realizations with our standard model (corresponding pink curves). model A results in a slightly larger $\sigma_z$ throughout (left). The difference between the standard model and model B is much more striking (right). Most importantly, the oldest stars in model B B have $\sigma_z$ about a factor of two smaller than when massive mergers are present. The highest radial velocity dispersion for the oldest stars (not shown) also drops form $\sim65$ to $\sim45$~km/s, which is reflected in the higher $V_\phi$-values. The lower velocity dispersions and rotational velocity difference between young and old stars is unlike the observations in the solar neighborhood. We can therefore argue that the MW thick disk is unlikely to have been formed through a quiescent disk evolution. An observational test must be devised to ascertain this possibility.  

\section{The chemical evolution model adopted in this work}
\label{sec:ap2}

We adopted a pure thin-disk chemical model (i.e., not the two-infall model of \citealt{chiappini97}), forming by continuous gas accretion (an exponential function of time -- see eq.~\ref{infall}). The galactic disk was divided into two-kpc-wide concentric rings that evolve independently without exchange of gas. We did not take into account radial gas flows or SN-driven winds (see Sec.~\ref{sec:tech}). Most of the observational constraints are confined to the solar vicinity (as summarized in Table~B.1).

\begin{table*}
\centering
\small
\textsc{{\bf Table B.1.} Main observational constraints for the solar neighborhood.}
\vspace{0.5cm}
\footnotesize
\label{obs}
\begin{tabular}{c c c}
\hline
Observable & Observed value & Reference\\
\hline
Surface densities of: & & \\
gas & 10--15  $M_\odot$ pc$^{-2}$ & \cite{kalberla09} \\
        & 7 $M_\odot$ pc$^{-2}$ & \cite{dickey93} \\
stars  & 35 $\pm$ 5 $M_\odot$ pc$^{-2}$ & \cite{gilmore95}\\
           & 35 $\pm$ 5 $M_\odot$ pc$^{-2}$ & \cite{flynn06} \\
stars (WDs + NSs) & 2--4 $M_\odot$ pc$^{-2}$ & \cite{mera98}\\
total & 56 $\pm$ 6 $M_\odot$ pc$^{-2}$ & \cite{holmberg04} \\
Star formation rate & 2--10 $M_\odot$ pc$^{-2}$ Gyr$^{-1}$ & \cite{guesten82}\\
                                  & 2.4 $M_\odot$ pc$^{-2}$ Gyr$^{-1}$ & \cite{fuchs09} \\
SN Ia rate & 0.43 -- 0.65 century$^{-1}$ & \cite{li11} \\
SN II rate & 1.27 -- 1.84 century$^{-1}$ & \cite{li11} \\
Infall rate & 0.3--1.5 $M_\odot$ pc$^{-2}$ Gyr$^{-1}$ & \cite{portinari98}\\
\hline
\end{tabular}
\end{table*}

An inside-out formation of the disk (in the sense of a more intense SFR in the inner regions - see Fig.~\ref{fig:sfh_sn}, middle panel) was simulated via a radial dependency of the gas accretion timescale (assumed linear for simplification), as follows

\begin{equation}\label{infall} 
A(r, t)= a(r)e^{-\frac{t}{\tau_{D(r)}}}, 
\end{equation} 
where A(r, t) is the gas infall rate and $\tau_D(r)$ is the timescale for the infalling gas onto the disk, given by (as in \citealt{chiappini01})
 
\begin{equation}
\tau_{D(r)} = 1.033 r - 1.27 \,\, \mbox{Gyr},
\end{equation}
for galactocentric distances $r\geq2$~kpc. 
 
The coefficient $a(r)$ in Eq.~\ref{infall} was chosen such as to reproduce the observed current \emph{total} surface mass density in the thin disk as a function of the galactocentric  distance given by

\begin{equation}
\Sigma(r) = \Sigma_{D}e^{-r/r_{D}},
\end{equation}
where $\Sigma_{D}$ is such that a total mass density of $\sim$50$~M_{\odot}$ pc$^{-2}$  is reached at the solar galactocentric distance (taken as 8 kpc), for a disk with a scale length $r_{D}= 3$~kpc (as shown in Fig.~\ref{fig:sfh_sn}, upper-left panel).

The SFR adopted here has the same formulation as in \cite{chiappini97}: 

\begin{equation}
\psi(R, t) \propto \Sigma_{gas}^{k}(R, t).
\end{equation}
The two free parameters are the gas surface density exponent (here taken as equal to 1.5) and the star formation efficiency, which we assumed is dependent on the total surface density as in \cite{chiappini97}.

We adopted the \cite{scalo86} IMF, constant in time and space. The stellar lifetimes, $\tau_{m}$, were taken from \cite{maeder89} as in our previous works. The impact of different assumptions on both IMF and stellar lifetimes on our chemical evolution results was discussed in detail in \cite{romano05}.

Type Ia SNe are the main contributors to the iron in the disk of our MW. The timescale on which their contribution becomes important is strongly dependent on the assumed SN progenitor model. In the so-called single degenerate scenario, a C-O white dwarf (WD) in a close binary system explodes due to gas accretion from a companion in a binary system. Another proposed scenario for Type Ia SNe is the so-called double degenerate case, where two WDs merge, thus triggering the explosion (see \citealt{matteucci09} and references therein). \cite{matteucci09} has recently shown that these two different SN Ia scenarios result in only negligible differences in the predicted [O/Fe] vs [Fe/H] relation in the MW. In the present work we adopted the single degenerate scenario. The Type Ia SN rate was computed following \cite{greggio83} and \cite{matteucci86} and is expressed as (see more details in \citealt{matteucci01})

\begin{equation}
R_{SNe Ia} = A_{Ia}\int\limits^{M_{BM}}_{M_{Bm}}\phi(M_{B})\left[
  \int\limits^{0.5}_{\mu_{m}}f(\mu)\psi(t-\tau_{M_{2}})d\mu \right] dM_{B},
\end{equation}
where $M_{2}$ is the mass of the secondary, $M_{B}$ is the total mass of the binary system, $\mu=M_{2}/M_{B}$, $\mu_{m}=max\left[M_{2}(t)/M_{B},(M_{B}-0.5M_{BM})/M_{B}\right]$, $M_{Bm}= 3 M_{\odot}$, $M_{BM}= 16 M_{\odot}$. The IMF is represented by $\phi(M_{B})$ and refers to the total mass of the binary system for the computation of the Type Ia SN rate, $f(\mu)$ is the distribution function for the mass fraction of the secondary,
\begin{equation}
f(\mu)=2^{1+\gamma}(1+\gamma)\mu^{\gamma},  
\end{equation}
with $\gamma=2$; $A_{Ia}$ is the fraction of systems in the appropriate mass range, which can give rise to Type Ia SN events. This quantity was fixed to 0.05 by reproducing the observed Type Ia SN rate at the present time (see Table~B.1). Note that in the case of the Type Ia SNe the so-called production matrix is indicated with $Q^{SNIa}_{mi}$ because of its different nucleosynthesis contribution (for details see \citealt{matteucci01}).

\begin{figure}
\centering
\includegraphics[width=8cm]{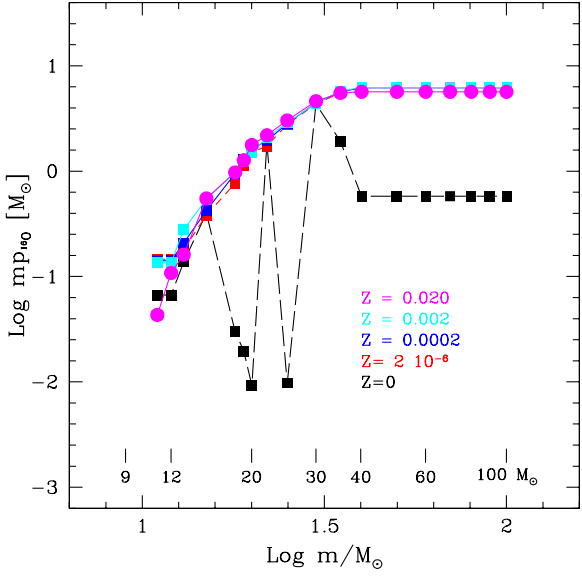}
\caption{oxygen stellar yields for massive stars from \cite{woosley95}. Here we adopt the solar yields (magenta circles). A stronger metallicity dependency of the oxygen yields is expected in models taking mass-loss into account (e.g., \citealt{meynet02}. The impact of different stellar yields is beyond the scope of the present paper and will be studied in future work.}
\label{fig:yieldsO}
\end{figure}

The equation below describes the time evolution of the gas mass fraction in the form of an element $i$, $G_{i}$ (see \citealt{matteucci01}):

\begin{displaymath}
\dot{G_{i}}(r,t)=-\psi(r,t)X_{i}(r,t) \\
\end{displaymath}
\begin{displaymath}
+ \int\limits^{M_{Bm}}_{M_{L}}\psi(r,t-\tau_{m})Q_{mi}(t-\tau_{m})\phi(m)dm \\
\end{displaymath}
\begin{displaymath}
+ A_{Ia}\int\limits^{M_{BM}}_{M_{Bm}}\phi(M_{B})\cdot\left[\int\limits_{\mu_{m}}^{0.5}f(\mu)\psi(r,t-\tau_{m2})Q^{SNIa}_{mi}(t-\tau_{m2})d\mu\right]dM_{B} \\
\end{displaymath}
\begin{displaymath}
+ (1-A_{Ia})\int\limits^{M_{BM}}_{M_{Bm}}\psi(r,t-\tau_{m})Q_{mi}(t-\tau_{m})\phi(m)dm \\
\end{displaymath}
\begin{displaymath}
+ \int\limits^{M_{U}}_{M_{BM}}\psi(r,t-\tau_{m})Q_{mi}(t-\tau_{m})\phi(m)dm \\
\end{displaymath}
\begin{equation}
+  X_{A_{i}}A(r,t),
\label{evol}
\end{equation}
 where $G_{i}(r,t)=[\Sigma_g(r,t) X_i(r,t)]/\Sigma_T(r)$, $\Sigma_g(r,t)$ is the surface gas density, and $\Sigma_T(r)$ is the present-time total surface mass density.  $X_{A_{i}}$ are the abundances in the infalling material, which here was taken as primordial. $X_{i}(r,t)$ is the abundance by mass of the element $i$ and $Q_{mi}$ indicates the fraction of mass restored by a star of mass $m$ in form of the element $i$, the production matrix (see \citealt{matteucci01} for more details). We indicated with $M_{L}$ the lightest mass that contributes to the chemical enrichment and set it to $0.8M_{\odot}$; the upper mass limit, $M_{U}$, was set to $100M_{\odot}$.

We adopted the stellar yields of \cite{woosley95} for the solar case, assuming no dependency on metallicity, for the contribution of massive stars. The adopted yields for oxygen are shown in Fig.~\ref{fig:yieldsO}. As it can be seen in the figure, for the models computed by \cite{woosley95}, an important difference is seen only at zero metallicities, which did not play a role in the present work (where we focused on the disk, and not the halo). In the case of iron, an important contribution comes also from Type Ia SNe. Here we adopted the stellar yields of \cite{iwamoto99}.

Finally, in Fig.~\ref{fig:gradO} we show the resulting oxygen abundance variation with radius at the present time in our model, compared to a sample of Cepheids whose distances are well known (see \citealt{cescutti07} and references therein). Fitting a line in the range $4<r<15$~kpc, we estimated a gradient of $\sim-0.04$~dex/kpc for the model. Given the young ages of the data ($\sim200$~Myr), these are expected to trace the gas gradient in the disk today, justifying the comparison in the figure.

\begin{figure}
\centering
\includegraphics[width=8.5cm]{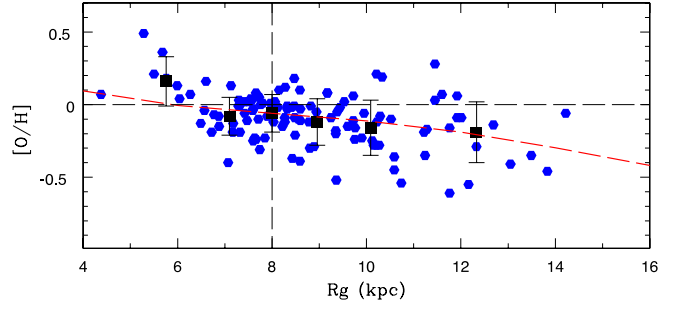}
\caption{Abundances of O as a function of galactocentric distance. The blue dots are the data by \cite{andrievsky02a, andrievsky02b, andrievsky02c, andrievsky04}. The black squares represent the mean values inside each bin and the error bars are the associated standard deviations. The red dashed line shows the prediction of our model, normalized to the mean value of the Cepheids at the galactocentric distance of the Sun. Fitting a line in the range $4<r<15$~kpc, we estimated a gradient of $\sim-0.04$~dex/kpc for the model, which matches the mean data points well.}
\label{fig:gradO}
\end{figure}

\end{appendix}

\end{document}